\documentclass[aps, a4paper,superscriptaddress, nofootinbib, showpacs, twocolumn, 10pt]{revtex4}
\usepackage{amsmath,amssymb, bm}
\usepackage{dsfont}
\usepackage{epsfig}
\usepackage{graphicx}
\usepackage{slashed}
\usepackage{color}
\usepackage{subfigure} 
\newcommand{\be}{\begin{equation}}
\newcommand{\ee}{\end{equation}}
\newcommand{\wt}{\widetilde}
\newcommand{\beq}{\begin{equation}}
\newcommand{\eeq}{\end{equation}}
\newcommand{\bea}{\begin{eqnarray}}
\newcommand{\eea}{\end{eqnarray}}
\newcommand{\ba}{\begin{align}}
\newcommand{\ea}{\end{align}}
\newcommand{\bfig}{\begin{figure}}
\newcommand{\efig}{\end{figure}}

\newcommand{\gev}{\, \text{GeV}}

\newcommand{\wh}{\widehat}
\newcommand{\nn}{\nonumber}

\begin{document}
~\vspace{1cm}
\title{
Expansions of $\tau$ hadronic spectral function moments in a nonpower
QCD perturbation theory with tamed large-order behavior}

\author{Gauhar Abbas}
\affiliation{The Institute of Mathematical Sciences, C.I.T.Campus,
Taramani, Chennai 600 113, India}
\author{B.Ananthanarayan}
\affiliation{Centre for High Energy Physics,
Indian Institute of Science, Bangalore 560 012, India}
\author{Irinel Caprini}
\affiliation{Horia Hulubei National Institute for Physics and Nuclear Engineering,
P.O. Box MG-6, 077125 Bucharest-Magurele, Romania}
\author{Jan Fischer}
\affiliation{Institute of Physics, Academy of Sciences of the Czech Republic, 
CZ-182 21  Prague 8, Czech Republic}
 
\begin{abstract}  
The moments of the hadronic spectral functions are of interest
for the extraction of the strong coupling  $\alpha_s$ and other QCD parameters 
from the hadronic decays of the $\tau$
lepton.  Motivated by the recent analyses of
a large class of moments in the standard fixed-order and
contour-improved perturbation theories, 
we consider the perturbative behavior of these moments in the framework of a QCD nonpower perturbation theory, 
defined by the technique of series acceleration by conformal mappings,
 which  simultaneously  implements renormalization-group 
summation and has a tame large-order behavior.  Two recently proposed 
models of the Adler function are employed to generate the higher order 
coefficients of the perturbation series and to predict the exact 
values of the moments, required for testing the properties of the perturbative expansions.
 We show  that the contour-improved nonpower perturbation theories and  the 
renormalization-group-summed nonpower perturbation theories  have very good convergence properties 
for a large class of moments of the so-called "reference model", including moments that are poorly described by the 
standard expansions. The results provide additional
support for the plausibility of the description of the Adler function 
in terms of a small number of dominant renormalons.

\end{abstract}
\maketitle

\section{Introduction} 
The strong coupling  $\alpha_s$ is a fundamental parameter  
whose determination is crucial for the low- and high-energy precision predictions of the standard model (SM).
A variety of sources exist for an accurate determination of this
quantity at different scales \cite{RPP,Altarelli2013,Pich2013}.  
The hadronic decays of the $\tau$ lepton allow for one of the most
precise determinations of the strong coupling and also
provide a beautiful experimental test of the predicted 
QCD running \cite{RPP,Pich2013}.  Indeed, the recent calculation 
of the QCD Adler function to five loops in massless QCD \cite{BCK08} 
motivated a large number of new determinations of $\alpha_s$ from 
these processes \cite{Davier2008,MaYa,BeJa,Pich_Muenchen,Pich_Manchester,Beneke_Muenchen,CaFi2009,CaFi_RJP,
CaFi_Manchester,CaFi2011,Abbas:2012,AACF,DV,Boito_update,CaSing,AACSing}.  It may however 
be noted that $\tau$ decays involve the strong coupling 
at a rather low scale, where the theoretical ambiguities 
inherent to perturbative QCD are expected to be large. 
An important ambiguity  is related to the prescription chosen for 
implementing renormalization-group invariance \cite{dLP1,Pivovarov:1991rh,dLP2,BeJa,Pich_Manchester}. Another serious problem 
is related to the fact that  the coefficients of the perturbative series 
of the Adler function in QCD display a factorial growth, {\em i.e.} 
the series has a vanishing radius of convergence \cite{ tHooft, Muell, Broa,Bene, Beneke}.  
These two problems are in fact related:  in particular, 
the inclusion of additional terms  in the expansion does not reduce, 
but on the contrary increases the dependence of the results on 
the renormalization-group prescription.  The nonperturbative 
power corrections and the effects of 
what is known as quark-hadron duality violation (DV),  
{\em i.e.} the breakdown of the operator product expansion  
near the timelike axis in the complex energy plane, 
generate additional uncertainties. The effects of these ambiguities 
are important especially at the  low scale  $M_\tau$, where the coupling 
is relatively large. The differences between the specific ways  
of treating them represent the main source of theoretical error 
in the extraction of  $\alpha_s(M_\tau^2)$.

The $\tau$ hadronic width is a good observable for the determination of 
the strong coupling, since it  receives small contributions from 
the power corrections and DV.  Various other moments 
have been also used in the past for the extraction of the strong coupling. 
Depending on the structure of the relevant weight, some moments
may receive larger contributions from the nonperturbative condensates and terms
involving DV, allowing the simultaneous extraction from data of these quantities
 and the strong coupling. 
The most comprehensive analysis to date, reported in \cite{Boito_update}, 
attempted to include DV in a combined fit of several moments, 
which in particular lead to a substantial increase in the error of the  
nonperturbative contributions. To improve such analyses, however, also the properties 
of the perturbative expansions of the moments must be carefully examined. 

Recently, the perturbative expansions of  a large class of spectral 
function moments have been discussed, under different assumptions 
for the large-order behavior of the Adler 
function \cite{BBJ}. 
This work extends the investigation of the hadronic width
 within  two standard QCD perturbative 
expansions, the fixed-order and the contour-improved perturbation theories
(FOPT and CIPT), to  moments defined by more general weights.
One of the important conclusions of \cite{BBJ} (and of the further study reported in \cite{Boito2013}) is that some moments 
that are commonly employed in $\alpha_s$ determinations from $\tau$ decays 
should be avoided because of their perturbative instability. We emphasize however that this refers
to the standard expansions, FOPT and CIPT. As we shall show in this paper, improved expansions with no
perturbative instability can be defined. 
 
It has been recently pointed out \cite{Abbas:2012,AACF,CaSing,AACSing} that an alternative to FOPT and CIPT,
which is placed somewhat between the two, but in practice is closer to CIPT, 
is one that sums the leading logarithms thereby accounting for
the renormalization-group invariance.  In Refs. \cite{AACF,CaSing,AACSing} we called this approach 
``renormalization-group-summed perturbation theory" (RGSPT).  In the present
work, we investigate the moments considered
in  \cite{BBJ} also in the frame of  RGSPT.  More significantly,
in the present paper we investigate the moments also
in the frame of a novel formulation of  QCD perturbation theory, 
defined some time ago in \cite{CaFi1998,CaFi2000,CaFi2001} starting 
from the divergent character of the standard series. The method uses 
the idea of series acceleration by means of a conformal mapping \cite{CiFi}, 
applied to the Borel plane of QCD correlators. In the new formulation,
 the standard powers of the coupling are replaced 
by  new expansion functions which are singular at the origin of the 
coupling plane  and have divergent perturbative  expansions,  
resembling thereby the expanded function  itself. To emphasize this essential feature, we named the new perturbation framework 
as ``nonpower perturbation theory`` (NPPT)  \cite{CaFi2011,CaSing,AACSing}.\footnote{We mention here that a different type of nonpower expansion, called ``analytic perturbation 
theory", which is not based on the idea of optimal conformal mapping but exploits the dispersion relations satisfied by the QCD correlators, has been proposed in \cite{Shirkov}.} Detailed 
studies
of the Adler function in the frame of NPPT \cite{CaFi2009}-\cite{AACF},
show that the best version is obtained by simultaneously
implementing renormalization-group invariance and the 
available knowledge about the divergent pattern of the series at large-orders. 
These optimized expansions were denoted as ``contour-improved nonpower perturbation theory" (CINPPT) 
and ``renormalization-group-summed nonpower perturbation theory" 
(RGSNPPT), respectively \cite{CaSing,AACSing}. 

Previous studies of the new expansions were focused on the extraction of  $\alpha_s$ 
from the total hadronic width, which involves  a particular moment 
of the spectral function.  We now generalize the investigation to 
the class of moments considered in \cite{BBJ}. The main aim of the work 
is to check whether the good convergence properties of CINPPT 
and RGSNPPT, demonstrated in the case of the hadronic width, 
remain valid also for the more general class of weights 
discussed in \cite{BBJ}.  

The scheme of this article is as follows:
In Sec. \ref{sec:moments} we  recall 
the definition of the spectral function moments,  
and specify the class of
moments investigated in \cite{BBJ}, that we consider also here.
We then briefly review in Sec. \ref{sec:pt} the standard 
perturbative expansions of the Adler function in massless QCD. 
In Sec.  \ref{sec:borel} we discuss the large-order behavior 
encoded in the singularities of the Borel transform. 
Here we point out the essential features of the Borel transform
in the three schemes, namely FOPT, CIPT and RGSPT.
In Sec. \ref{sec:nppt}, using the technique of ``optimal conformal mapping" (OCM) and ``singularity softening" for convergence acceleration, we define, for each RG prescription, 
a class of new, nonpower expansions, where the powers of the coupling are replaced by more general functions. 
The models proposed in \cite{BeJa, BBJ} for the physical Adler function, denoted as the reference model (RM) and the
alternative model (AM), are 
briefly reviewed in Sec. \ref{sec:models}.  These models are used to compute the higher-order perturbative coefficients, as well as the exact value and the ambiguity of the moments.
 Our results on the perturbative expansions of the moments are presented
in Sec. \ref{sec:results}, which we  split into
several subsections to facilitate the discussion: we first consider  moments defined by integrals  up to $s_0=M_\tau^2$, expanded in the frame of CINPPT and RGSNPPT based on the OCM. In the 
next subsection we explore a larger class of expansions based on different softening factors and different conformal mappings, and in the last
 subsection we consider moments defined by integrals up to an $s_0$  lower than $M_\tau^2$.  
Section \ref{sec:conc} contains discussions and conclusions.

\section{ Moments of the spectral functions}\label{sec:moments}
We consider the moments of the spectral function ${\rm Im} \Pi^{(0+1)}(s)$ 
defined as \cite{BBJ}
\beq\label{eq:delwi}
M_{w_i}(s_0)= 
\frac{2}{\pi}\int\limits_0^{s_0} w_i(s/s_0) \,{\rm Im} \Pi^{(0+1)}(s)\,ds,
\eeq 
where $s_0\leq M_\tau^2$ and  $w_i(x)$ are arbitrary nonnegative weights. 
We are interested in the perturbative  contribution to $M_{w_i}$ dependent on $\alpha_s$,
 denoted as $\delta^{(0)}_{w_i}$, obtained by subtracting 
from (\ref{eq:delwi}) the perturbative tree values $\delta^{\rm tree}_{w_i}(s_0)$.
We adopt the set of weights $w_i(x)$, $i=1,\,17$ investigated in \cite{BBJ}. 
For the purposes to follow, we need to define in terms of the $w_i$,
the corresponding
\beq\label{eq:Wi}
W_i(x)=2\int_x^1 dz w_i(z).
\eeq
For completeness, we list in Table \ref{tab:Wi}  
the functions $W_i(x)$ for the weights $w_i(x)$ adopted in \cite{BBJ}. 
We recall that $i=12$ gives the kinematical weight $w_\tau$ relevant for 
the hadronic decay width. According to the terminology of \cite{BBJ}, 
the first class in  Table \ref{tab:Wi} contains functions $W_i(x)$ 
generated from weights $w_i(x)$ equal 
to the monomials $x^{i-1}$, the second class 
contains moments generated by  ``pinched" weights ({\em i.e.} weights 
that vanish at $x=1$)  and include a ``1" term, 
and the third class contains  ``pinched"  
weights without a ``1" term, respectively. 
Some of the moments listed in Table \ref{tab:Wi} of
the first and second classes were investigated in Ref.~\cite{CaFi2011}.

\begin{table}
\centering
\begin{tabular}{l l  }\toprule
 $i$ ~~~~~ & $W_i(x)$ \\ \hline
  1 ~~~~~ & $2(1-x)$     \\
  2 ~~~~~ & $1-x^2$  \\
  3 ~~~~~ & $\frac{2}{3}(1-x^3)$  \\
  4 ~~~~~ & $\frac{1}{2}(1-x^4)$\\
  5 ~~~~~ & $\frac{2}{5}(1-x^5)$ \\ [0.1cm]\hline
  6 ~~~~~ & $ (1-x)^2$\\ 
  7 ~~~~~ & $ \frac{2}{3} (1 - x)^2 (2 + x) $\\
  8  ~~~~~& $ \frac{1}{2}(3-4 x +x^4)$\\
  9 ~~~~~ & $ \frac{1}{4}(1-x)^3(3+x)$  \\
 10 ~~~~~ & $ \frac{2}{3}(1-x)^3 $ \\
 11 ~~~~~ & $\frac{1}{2} (1-x)^4$\\
 12 ~~~~~ & $(1-x)^3(1+x)$  \\ 
 13 ~~~~~& $ \frac{1}{10}(1-x)^4(7+8 x)$ \\[0.1cm] \hline
 14 ~~~~~ & $ \frac{1}{6}(1-x)^3 (1+3 x)$ \\
 15 ~~~~~& $ \frac{1}{6} (1 - x)^4 (1 + 2 x)^2$ \\
 16 ~~~~~ & $ \frac{1}{210}(1-x)^4( 13 + 52 x + 130 x^2 + 120 x^3)$  \\
 17 ~~~~~ & $\frac{1}{70}(1-x)^4(2 + 8 x + 20 x^2 + 40 x^3 + 35 x^4 )$  \\[0.1cm] \toprule
\end{tabular}
\caption{Functions $W_i(x)$ defined in (\ref{eq:Wi}) for the weights $w_i(x)$ listed in Table 2 of \cite{BBJ}. }
\label{tab:Wi}
\end{table}

The analytic properties of the 
polarization function and the Cauchy theorem allow one to write  
equivalently (\ref{eq:delwi})  as an integral along a contour in the complex
$s$ plane, chosen for convenience to be the circle $|s|=s_0$. After an integration by parts, the perturbative contribution
 $\delta^{(0)}_{w_i}$ can be written as
\begin{equation}\label{eq:del0}
\delta^{(0)}_{w_i}(s_0)= \frac{1}{2\pi i} \!\!\oint\limits_{|s|=s_0}\!\! \frac{d s}{s} 
W_i(s/s_0) \widehat{D}_{\rm pert}(s),
\end{equation}
where the weights $W_i(x)$ are defined in (\ref{eq:Wi}) and $\widehat{D}_{\rm pert}$ is the perturbative part 
of the reduced Adler function 
\beq\label{eq:Dhat}
\widehat{D}(s)\equiv - s\, {\rm d}\Pi^{(1+0)}(s)/{\rm d}s -1.\eeq 
This sets the stage for the computation of the moments of interest
in this work.

\section{Renormalization-group summation: FOPT, CIPT and RGSPT}\label{sec:pt}
 In our notation the standard perturbative expansion of the Adler function in a definite renormalization scheme, denoted usually as FOPT \cite{BeJa},  
is written as
\beq \label{eq:fopt}
\widehat{D}_{\rm FOPT}(s) = \sum\limits_{n\ge 1} (a_s(\mu^2))^n [c_{n,1}+
\sum\limits_{k=2}^{n} k  c_{n,k} \left(\ln\frac{-s}{\mu^2}\right)^{k-1}],\eeq
where $a_s(\mu^2)=\alpha_s(\mu^2)/\pi$.  In (\ref{eq:fopt}) the renormalization scale $\mu^2$ is  chosen close to $s_0$,  the leading coefficients $c_{n,1}$  are computed from Feynman diagrams,
and  $c_{n,k}$ for $2 \leq k \leq n $ depend on $c_{n,1}$ and  the perturbative coefficients $\beta_k$  of the renormalization-group (RG) $\beta$ function, which are known at present 
to four loops \cite{LaRi,Czakon}.  In the  ${\overline{\rm MS}}$ scheme for $n_f=3$ flavors  the coefficients $c_{n,1}$  calculated up to fourth order  (cf. \cite{BCK08} and 
references therein) are:
\beq\label{eq:cn1} c_{1,1}=1,~ c_{2,1}= 1.64,~ c_{3,1}=6.371,~ c_{4,1}=49.079.\eeq

By setting $\mu^2=-s$ in the expansion (\ref{eq:fopt}), one obtains the CIPT expansion of the Adler function \cite{Pivovarov:1991rh, dLP1, dLP2, Pich_Manchester}:
\beq \label{eq:cipt}\widehat{D}_{\rm CIPT}(s)= \sum\limits_{n\ge 1} c_{n,1} \,(a_s(-s))^n, ~~~~~ \eeq
where the running coupling  $a_s(-s)$  is determined by solving the RG equation 
\beq\label{RGE}
s \frac{da_s (-s)}{ds} = \beta (a_s).
\eeq 
For the evaluation of the  integral (\ref{eq:del0}), this equation is solved numerically in an iterative way along the  contour $|s|=s_0$,  starting with the input value $a_s(M_\tau^2)$ at $s=-M_\tau^2 $.

 The properties of the above expansions, in particular their convergence  and the behavior in the complex energy plane, have been examined critically in several recent 
papers \cite{Davier2008, BeJa,Pich_Manchester, Beneke_Muenchen,BBJ}, where arguments in favor of one or another expansion have been given.

We mention also another prescription, proposed in \cite{Ahmady1,Ahmady2} for timelike observables and applied in  \cite{Abbas:2012,AACF} to Adler function in the complex energy plane. It 
generalizes the summation of leading logarithms to nonleading logs, by summing all the terms available from RG invariance. We refer to it as RGSPT. 
 It can be shown \cite{AACF} that the perturbative expansion 
(\ref{eq:fopt}) can be written as 
\beq \label{eq:rgspt}\wh D_{\rm RGSPT}(s) = \sum_{n\ge 1} \,(\wt a_s(-s))^n [c_{n,1}+ \sum_{j=1}^{n-1} c_{j,1} d_{n,j}(y)], \eeq
where 
\beq\label{eq:atilde}
\wt a_s(-s)=\frac{a_s(\mu^2)}{1+\beta_0 a_s(\mu^2) \ln(-s/\mu^2)}
\eeq
  is the solution of the RG equation (\ref{RGE})  to one loop, and 
 $d_{n,j}(y)$ are  calculable functions depending on the variable $y\equiv 1+\beta_0 a_s(\mu^2) \ln(-s/\mu^2)$ and the coefficients  $\beta_j$. These functions are shown to vanish 
for $y=1$ or in the limit $\beta_j=0$, $j\ge 1$. They have analytically closed, but quite lengthy expressions, given in 
\cite{AACF}. 
 As an effective series in powers of the one-loop running coupling, with coefficients that depend still on the coupling at a fixed scale,  and also on the nonleading $\beta_j$, 
the expansion (\ref{eq:rgspt}) appears to be placed ``in-between" FOPT and CIPT: it resembles FOPT as it contains only analytical closed expressions, but  makes a summation of higher terms known from renormalization-group invariance, like CIPT. Actually, in practice, since the running of $\alpha_s$ in QCD is
largely dominated by $\beta_0$, the RGS expansion and CIPT are very
similar. This feature is confirmed numerically, as discussed in detail in  \cite{Abbas:2012}. 
\section{Large-order behavior and the Borel transform}\label{sec:borel}

From special classes of Feynman diagrams  it is known  that the perturbative coefficients $c_{n,1}$ display a factorial increase, $c_{n,1}\sim n!$, so the perturbative expansions 
written above are divergent series \cite{Muell, Broa, Bene,Beneke}. This property follows also indirectly from the  arguments given in \cite{tHooft}, which infer that the expanded 
amplitude, viewed as a function of the coupling $\alpha_s$, is singular at $\alpha_s=0$. The divergent series in field theory are often interpreted as asymptotic 
series  \cite{Dyson,Muell,Beneke}. 

 The large-order behavior of the CIPT series (\ref{eq:cipt}) is encoded in the properties of the Borel transform $B(u)$, defined by the expansion\footnote{  For consistency with our  
subsequent notations,  this Borel transform should have the index ``CI". However, we prefer the standard notation $B(u)$, which is used by most authors.}
\beq\label{eq:B}
B(u)= \sum_{n=0}^\infty c_{n+1,1}\, \frac{u^n}{\beta_0^n \,n!}.
\eeq
The original function $\wh D_{\rm CIPT}(s)$  is recovered from $B(u)$ by a Laplace-Borel integral. Actually, in the present case $ B(u)$ is known to have singularities on the real 
axis of the $u$ plane, more precisely along the lines $u\ge 2$ (infrared  renormalons)  and $u\le -1$ (ultraviolet  renormalons) \cite{Beneke}, so the integral requires a prescription. 
We adopt  the principal value ({\rm PV}) prescription \cite{Muell,Beneke,BeJa}  
\beq\label{eq:LaplaceCI}
 \wh D_{\rm CIPT}(s)=\frac{1}{\beta_0}\,{\rm PV} \,\int\limits_0^\infty  
\exp{\left(\frac{-u}{\beta_0 a_s(-s)}\right)
} \, B(u)\, {\rm d} u\,,
\eeq
which is preferred from the point of view of momentum-plane analyticity \cite{CaNe}.

 Similarly, one defines the Borel transforms $B_{\rm FO}(u,s)$ and $B_{\rm RGS}(u, y)$  of the FOPT and RGSPT expansions, (\ref{eq:fopt}) and (\ref{eq:rgspt}) respectively, which can be 
written as \cite{AACF}: 
\beq\label{BFO}
B_{\rm FO}(u,s)=B(u)+ \sum_{n=0}^\infty  \frac{u^n}{\beta_0^n \,n!} \sum\limits_{k=2}^{n+1} k c_{n+1,k}\left(\ln\frac{-s}{M_\tau^2}\right)^{k-1},
\eeq
\beq\label{BRGS}
B_{\rm RGS}(u, y)=B(u) + \sum_{n=0}^\infty\frac{ u^n} {\beta_0^n n!} \sum_{j=1}^{n} c_{j,1} d_{n+1,j}(y).
\eeq
The functions $\wh D_{\rm FOPT}(s)$ and $\wh D_{\rm RGS}(s)$ are recovered from their Borel transforms  by Laplace-Borel integrals similar to (\ref{eq:LaplaceCI}):
\beq\label{eq:LaplaceFO}
 \wh D_{\rm FOPT}(s)=\frac{1}{\beta_0}{\rm PV} \,\int\limits_0^\infty  
\exp{\left(\frac{-u}{\beta_0 a_s(s_0)}\right)
} B_{\rm FO}(u,s) {\rm d} u,
\eeq
\beq\label{eq:LaplaceRGS}
 \wh D_{\rm RGSPT}(s)=\frac{1}{\beta_0}{\rm PV}\int\limits_0^\infty  
\exp{\left(\frac{-u}{\beta_0 \tilde a_s(-s)}\right)
} B_{\rm RGS}(u, y){\rm d} u .
\eeq

It is important to recall that not only the location, but also the nature of the leading singularities of $B(u)$ is known. Namely, near the points $u=-1$ and $u=2$ $B(u)$ behaves as 
\beq\label{eq:Bthr}
B(u) \sim (1+u)^{-\gamma_{1}},  \quad B(u)  \sim (1-u/2)^{-\gamma_{2}}, \eeq
where  $\gamma_1 = 1.21$ and  $\gamma_2 = 2.58$  \cite{Muell,Beneke,BBK,BeJa, BBJ}. As argued in \cite{Muell,AACF}, the leading singularities in the $u$ planeof the Borel transforms $B_{\rm FO}(u,s)$ and $B_{\rm RGS}(u, y)$ have the same 
positions and nature as those of $B(u)$.

Starting from the divergent character of the standard perturbative series in QCD, the need of a new  perturbation theory  was advocated in \cite{CaFi1998}.
Since the powers of $a_s$ are holomorphic, while the function $\wh D_{\rm pert}$ is expected to be singular at the expansion point $a_s=0$, no finite-order standard 
perturbative approximant can share this singularity
with the expanded function: singularities can emerge only from the infinite 
series as a whole, which is not defined unambiguously since the perturbation series is divergent.

As discussed in \cite{CaFi2011, CaSing}, a perturbation series would be more instructive if the
finite-order approximants could retain some information about the 
known singularities of the expanded function.  Such approximants would tell us more about the function
also from the numerical point of view. In the next section we shall review, following \cite{CaFi1998, CaFi2011, AACF, CaSing}, the properties of these improved expansions 
based on the technique of series acceleration by the conformal mappings of the complex plane.
\section{Nonpower perturbative expansions}\label{sec:nppt}
As discussed in Ref. \cite{CaFi2011}, the method of conformal mappings is  not applicable to the (formal) perturbative series in powers of  
$a_s$, because the expanded correlators  are singular 
at the point of expansion, $a_s=0$. However, the method
can be applied  in the Borel  plane, where a holomorphy domain around the origin $u=0$ is known to exist.

The starting point in the derivation is the remark that the expansion (\ref{eq:B}) converges only in the disk $|u|<1$, whose boundary passes through the  
singularity of $B(u)$ closest to $u=0$. However, the function $B(u)$ is holomorphic in a larger domain, assumed in general to be the whole complex $u$ plane cut 
along the lines $u\ge 2$ and $u\le -1$. It would be useful to insert in (\ref{eq:LaplaceCI}) an expansion of $B(u)$  that is convergent also outside the disk $|u|<1$. Such an expansion 
is easily obtained: since the disk is the natural domain of convergence  for power series, it suffices to expand the function in powers of variables that perform the conformal mapping of 
a larger part of its holomorphy 
domain   onto a disk.  Intuitively, one expects that a larger  domain of convergence is related also to a better convergence rate. This expectation turns out to be correct: as shown a 
long time ago in \cite{CiFi}, the variable that maps the entire holomorphy  domain of the expanded function onto a disk  has the remarkable property that the expansion in powers of this 
variable has the fastest large-order 
convergence rate at all points inside the holomorphy domain (we assume here that the holomorphy domain 
is simply connected).  This mapping was called ``optimal conformal mapping"  for series expansions \cite{CiFi}. More detailed arguments are given in two lemmas  formulated 
and proved in \cite{CaFi_Manchester,CaFi2011}. For the Adler function 
in QCD, the optimal mapping $\wt w(u)$ and the corresponding perturbative expansion were defined and investigated in \cite{CaFi1998,CaFi2000,CaFi2001}. 

An additional improvement is obtained by exploiting the known
 behavior (\ref{eq:Bthr}) of $B(u)$ near the first singularities. 
If one multiplies $B(u)$ with a suitable factor  $S(u)$, which fully compensates, or at least ``softens" the dominant singularities, the expansions will have a more rapid convergence even at low orders \cite{SoSu}.  In fact,  a mild branch point, where the function vanishes instead of becoming infinite, is expected to influence the power expansions of the function only at larger orders. 
Hence, one can  expand the product $S(u) B(u)$ in powers of conformal mappings that account only for the nonleading, {\em i.e.} the more distant, singularities, and contain a residual ``mild" cut 
inside the convergence disks. In view of these remarks, it was useful to consider the  general class of conformal mappings  \cite{CaFi_RJP,CaFi_Manchester,CaFi2011,CaFi1998}:
\beq\label{eq:wjk}  w_{jk}\equiv \wt w_{jk}(u)=\frac{\sqrt{1+u/j}-\sqrt{1-u/k}}{\sqrt{1+u/j}+\sqrt{1-u/k}},\eeq
where $j, k$ are positive integers satisfying  $j\ge 1$ and $k \ge 2$.  The function $\wt w_{jk}(u)$ maps the $u$ plane cut along $ u\le -j$ and $u\ge k$ onto the disk $|w_{jk}|<1$ in 
the plane $w_{jk}\equiv \wt w_{jk}(u)$, such that $\wt w_{jk}(0)=0$, $\wt w_{jk}(-j)=-1$ and $\wt w_{jk}(k)=1$. The OCM defined above is $\wt w(u)\equiv \wt w_{12}(u)$, for which the entire holomorphy domain of the Borel transform, {\em i.e.} the $u$ plane cut along $u\ge 2$ and $u\le -1$, is mapped onto the 
interior of the unit circle in the plane $w_{12}\equiv \wt w_{12}(u)$. 

Using the above ideas, we consider the following expansion  \cite{CaFi_Manchester,CaFi_RJP, CaFi2011,CaFi1998}
\beq\label{eq:Bw}
 S(u) B(u) =\sum_{n\ge 0} c_{n, {\rm CI}}^{(jk)}\,  (\wt w_{jk}(u))^n.
\eeq
 In practice, this series is obtained by inserting in the product $S(u) B(u)$ the series (\ref{eq:B}) truncated at the order $N$, with $u$ replaced by
 the inverse $\tilde u_{jk}$  of (\ref{eq:wjk}). Then we expand the product in powers of $ \wt w_{jk}(u)$ and keep $N$ terms in the series.

As discussed in \cite{CaFi2009,CaFi2011}, unlike the OCM which is unique, the choice of  the softening factor $S(u)$, {\em i.e.} the implementation of the known nature of the first branch points, is  to a large extent arbitrary. For a large number of terms in the expansion (\ref{eq:Bw}),
the form of this factor should be irrelevant, but at low orders 
one prescription may be better than another. 

In Refs. \cite{CaFi2011, AACF} the factor $S(u)$ was  chosen as a 
simple expression of the expansion variable  $\wt w_{jk}(u)$ itself 
\beq\label{eq:Sjk}
 S(u)\equiv S_{jk}(u)=\left(1-\frac{\wt w_{jk}(u)}{\wt w_{jk}(-1)}\right)^{\gamma'_1} \left(1-\frac{\wt w_{jk}(u)}{\wt w_{jk}(2)}\right)^{\gamma'_2}, \eeq
where $ \gamma_j'$, $j=1,2$, are suitable exponents, given in \cite{CaFi2011}, defined such as to preserve the behavior (\ref{eq:Bthr}) of $B(u)$.
 This choice ensures a good convergence of the expansion (\ref{eq:Bw}),  as noted by  extensive  numerical calculations \cite{CaFi2011}. Of course, other choices are possible, for instance the simple  expression 
\beq\label{eq:Su}
 S(u)=(1+u)^{\gamma_1}(1-u/2)^{\gamma_2}.
\eeq
 The expansions 
(\ref{eq:Bw}) converge in a domain larger than the convergence  disk $|u|<1$ of the original series (\ref{eq:B}), and according to the lemmas proven in \cite{CaFi2011},  have a better 
convergence rate, in particular at points $u$ close to the origin, which are dominant in the  Laplace-Borel integral (\ref{eq:LaplaceCI}). 
The use of several conformal mappings and different softening factors  reduces the bias related to the implementation of the threshold behavior (\ref{eq:Bthr}), which is not 
unique, as we mentioned above.  As discussed in \cite{CaFi2011}, useful choices of the expansion variables are, besides the OCM $\wt w_{12}(u)$, the conformal mappings $\wt w_{13}(u)$,  $\wt w_{1\infty}(u)$ and $\wt w_{23}(u)$. 

 From (\ref{eq:LaplaceCI}) and (\ref{eq:Bw})  one
obtains the CINPPT \cite{CaFi2011}
\beq\label{eq:cinppt} \wh D_{\rm CINPPT} (s)= \sum\limits_{n \ge 0} c_{n, {\rm CI}}^{(jk)} \, {\cal W}^{jk}_{n, {\rm CINPPT}}(s),\eeq 
where the expansion functions are defined as
\beq\label{eq:Wnpci}  {\cal W}^{jk}_{n, {\rm CINPPT}}(s)=\frac{1}{\beta_0} {\rm PV} \int\limits_0^\infty\!{\rm e}^{-u/(\beta_0 a_s(-s))} \,  \frac{(\wt w_{jk}(u))^n}{S(u)}\,{\rm d} u. \eeq
Similarly, the ``fixed-order nonpower perturbation theory" (FONPPT) and the ``renormalization-group-summed nonpower perturbation theory" (RGSNPPT) are defined as \cite{CaFi2011, AACF}
\beq\label{eq:fonppt} \wh D_{\rm FONPPT}  (s) = \sum\limits_{n\ge 0} c_{n, {\rm FO}}^{(jk)}(s) \,{\cal W}^{jk}_{n, {\rm FONPPT}}(s_0),\eeq
\beq \label{eq:rgsnppt} \wh D_{\rm RGSNPPT}  (s) = \sum\limits_{n\ge 0} c_{n, {\rm RGS}}^{(jk)}(y) \, {\cal W}^{jk}_{n, {\rm RGSNPPT}}(s), \eeq
where the coefficients are obtained from expansions similar to (\ref{eq:Bw}) of the Borel transforms $B_{\rm FO}(u,s)$ and $B_{\rm RGS}(u,y)$, and the expansion functions ${\cal W}^{jk}_{n, {\rm FONPPT}}(s_0)$ and ${\cal W}^{jk}_{n, {\rm RGSNPPT}}(s)$ are obtained from (\ref{eq:Wnpci}) by replacing the running 
coupling $a_s(-s)$ in the exponent through the fixed-scale coupling $a_s(s_0)$ and the one-loop running coupling $\wt a_s(-s)$ defined in (\ref{eq:atilde}), respectively.

The properties of the expansions (\ref{eq:cinppt})-(\ref{eq:rgsnppt}) have been discussed in \cite{CaFi1998,CaFi2001,CaFi2011}. 
When reexpanded in powers of $a_s$, they  reproduce 
order by order the known perturbative coefficients calculated from Feynman
diagrams. On the other hand, 
 the expansion functions resemble the expanded function, being singular at   $a_s=0$ and having divergent series in powers of $a_s$. Therefore, the divergent pattern of the expansion of 
the QCD correlators in terms of these
new functions is expected to be tamer. This expectation is fully confirmed for the expansions that implement also RG summation, {\em i.e.} CINPPT and RGSNPPT, which give 
a very good description of suitable models of the Adler function in the complex $s$ plane \cite{CaFi2011, AACF}. By contrast, FONPPT  gives a very good description  near the spacelike axis, 
which gradually deteriorates for points closer to the timelike axis. As discussed in \cite{CaFi2009,CaFi2011}, this behavior is due to the large imaginary parts of the logarithms in the 
coefficients  $c_{n, {\rm FO}}^{(jk)}(s)$ near the timelike axis, which follow from (\ref{eq:fopt}) and  (\ref{BFO}). Therefore,  FONPPT describes well "pinched"  moments for which 
the weight function suppresses this region, but the description is not so good for other moments. For this reason we shall concentrate in this paper mainly on the   CINPPT 
and RGSNPPT frameworks, which simultaneously sum the large logarithms by RG invariance and tame the large-order behavior,  by accelerating the convergence through conformal mappings.   
We shall present some results obtained with FONPPT only to illustrate the statement made above.

\section{ Models 
for the Adler functions}\label{sec:models}
In order to test numerically the convergence properties
of the perturbative expansions,  a model for the higher-order coefficients of the Adler function, $c_{n,1}$ for $n>4$, is necessary. We follow the approach  
adopted recently in the literature \cite{BeJa, CaFi2009, CaFi2011, DeMa, BBJ}, in which the physical function is expressed 
in terms of a few dominant singularities in the Borel plane. Unfortunately, even in this rather limited class of models, considerable freedom  still exists: 
while the nature  of the leading singularities is known, the residues cannot be determined from theory and an ansatz must be adopted.  As 
discussed in \cite{Pich_Manchester, DeMa, BBJ}, depending on the assumed strength pattern of the dominant singularities, either FOPT or CIPT turns out to be the preferred scheme. 

In the RM proposed
in \cite{BeJa, BBJ},  the  Adler function $\wh D(s)$ is defined as the {\rm PV}-regulated Laplace-Borel integral
\beq\label{eq:pv}
\wh D(s)=\frac{1}{\beta_0}\,{\rm PV} \,\int\limits_0^\infty  
\exp{\left(\frac{-u}{\beta_0 a_s(-s)}\right)} \, B(u)\, {\rm d} u
\eeq
where the Borel transform $B(u) \equiv B_{\rm RM}(u)$ is parametrized in terms of a few ultraviolet (UV) and infrared (IR) renormalons, and a regular, polynomial part. In our notations, it reads
\beq\label{eq:BBJ}
\frac{B_{\rm RM}(u)}{\pi}=B_1^{\rm UV}(u) +  B_2^{\rm IR}(u) + B_3^{\rm IR}(u) +
d_0^{\rm PO} + d_1^{\rm PO} u, 
\eeq
with  the renormalons parametrized as \cite{BeJa,BBJ}
\bea\label{eq:BIRUV}
B_p^{\rm IR}(u)= 
\frac{d_p^{\rm IR}}{(p-u)^{\gamma_p}}\,
\Big[\, 1 + \wt b_1 (p-u) + \ldots \,\Big], \nonumber \\
B_p^{\rm UV}(u)=\frac{d_p^{\rm UV}}{(p+u)^{\bar\gamma_p}}\,
\Big[\, 1 + \bar b_1 (p+u)  +\ldots \,\Big].
\eea
The free parameters of the model, namely the residues $d_1^{\rm UV},\, d_2^{\rm IR}$ and  $d_3^{\rm IR}$ of the first renormalons and the coefficients $d_0^{\rm PO}, d_1^{\rm PO}$ of the polynomial in (\ref{eq:BBJ}),  were determined  by the requirement of reproducing 
the perturbative coefficients $c_{n,1}$ for $n\le 4$ from (\ref{eq:cn1}) 
and the estimate $c_{5,1}=283$. Their numerical values are \cite{BeJa, BBJ}:
\bea\label{eq:dBJ}
&&d_0^{\rm PO}=0.781, ~~~
d_1^{\rm PO}=7.66\times 10^{-3},\\
&& d_2^{\rm IR}=3.16,~~ d_3^{\rm IR}=-13.5,~~ d_1^{\rm UV}=  - 1.56 \times 10^{-2}. \nn
\eea
Then all the higher order coefficients $c_{n,1}$ are fixed and 
exhibit a factorial increase. Their numerical values  up to $n=18$ are listed in \cite{BeJa,CaFi2009}. 

This model is considered in \cite{BeJa,BBJ} as most natural from the point of view of the strengths of the leading singularities, as no residue is fixed {\em a priori} by hand. If this 
model is adopted, FOPT provides the preferred framework for implementing RG invariance of the spectral function moments \cite{BBJ}.

On the other hand, models with a smaller residue $d_2^{\rm IR}$ of the first IR renormalon are described better by  CIPT \cite{DeMa, CaFi2011, BBJ}. As an extreme case, in the 
AM considered in \cite{BBJ} the first IR renormalon at $u=2$ was removed by hand. Thus, the ``extreme" AM proposed in  \cite{BBJ} is defined by a 
Borel transform  $B(u)\equiv B_{\rm AM}(u)$ containing no singularity at $u=2$ and an additional singularity at $u=4$:
\bea\label{eq:altBBJ}
\frac{B_{\rm AM}(u)}{\pi}
 &=&B_1^{\rm UV}(u) +  B_3^{\rm IR}(u) + B_4^{\rm IR}(u)  
\nn\\
&+& d_0^{\rm PO} + d_1^{\rm PO} u.
\eea
The five parameters  found by matching the coefficients $c_{n,1}$ for 
$n\le 5$ are: 
\bea\label{eq:altdBJ}
&&d_0^{\rm PO}=2.15, ~~ d_1^{\rm PO}=4.01 \times 10^{-1},\\ 
&&d_3^{\rm IR}=66.18, ~~ d_4^{\rm IR}=-289.71, ~~d_1^{\rm UV}=-5.21 \times
10^{-3}. \nn
\eea
Intermediate models, where the IR renormalon at $u=2$ is present, but has a prescribed residue smaller (or larger) than the value in (\ref{eq:dBJ}), were discussed in \cite{DeMa, Pich_Muenchen,CaFi2011}.

  The  properties of the perturbative expansions of the quantities 
$\delta^{(0)}_{w_i}$ in the standard FOPT and CIPT, using the models
 described above, were discussed in detail in  \cite{BBJ,Boito2013}. 
The parameter $s_0$ was set equal to $M_\tau^2$ in \cite{BBJ}, while lower values 
of $s_0$ were investigated in \cite{Boito2013}.
 For each class of 
weights defined in Table \ref{tab:Wi}, specific features of the perturbative expansions were identified. 
A bad perturbative behavior was found for some moments, 
in particular from the last class in Table \ref{tab:Wi} 
and for weights $w_i(x)$ having a linear term in $x$.
\section{Results}\label{sec:results}

\begin{figure*}\begin{center}

\vspace{.5cm}
\subfigure[~$W_1$, reference model]{\includegraphics[width=.8\columnwidth,angle=0]{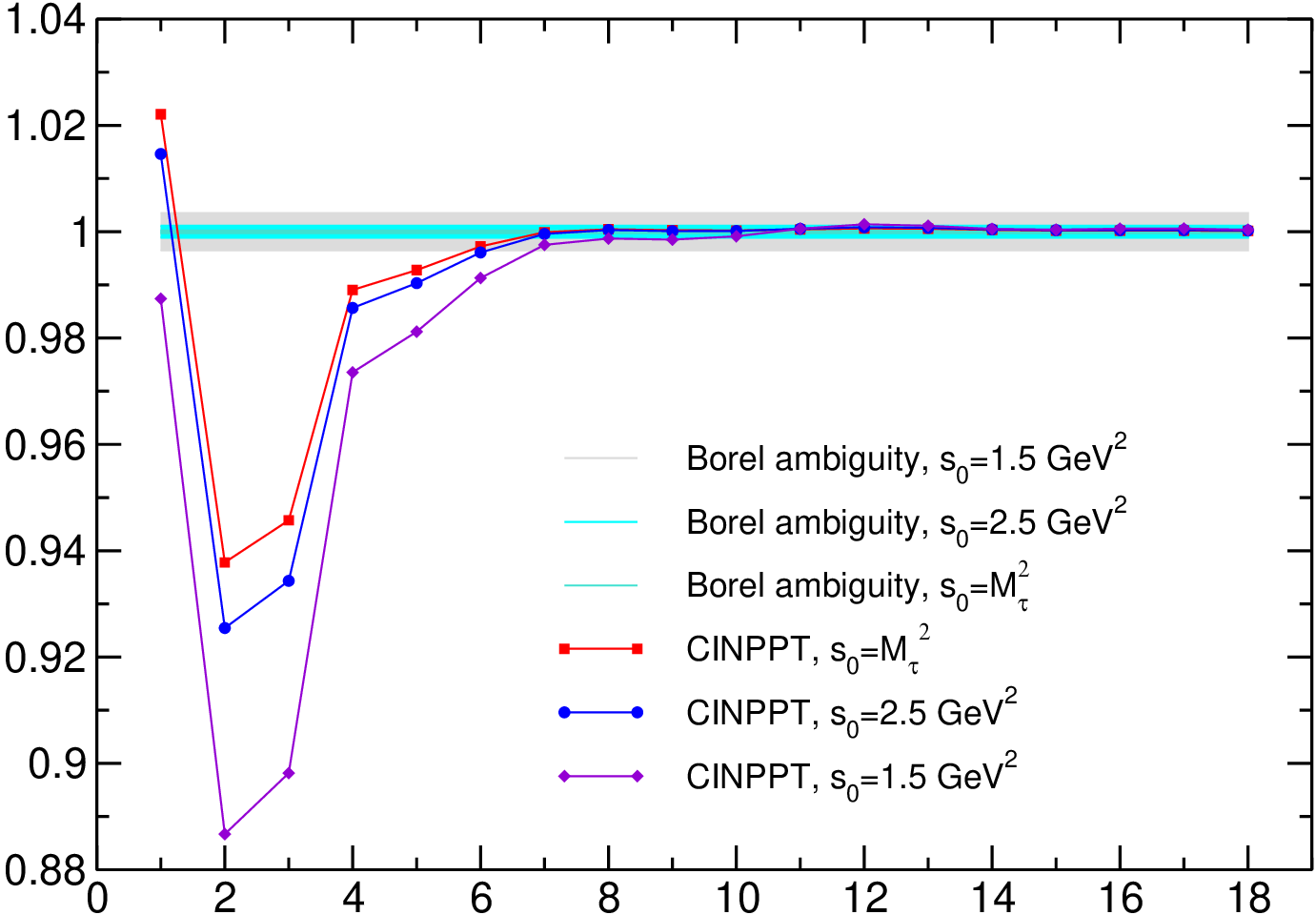}\label{w1RMs0}}\hspace{1cm}
\subfigure[~$W_{2}$, reference model]{\includegraphics[width=.8\columnwidth,angle=0]{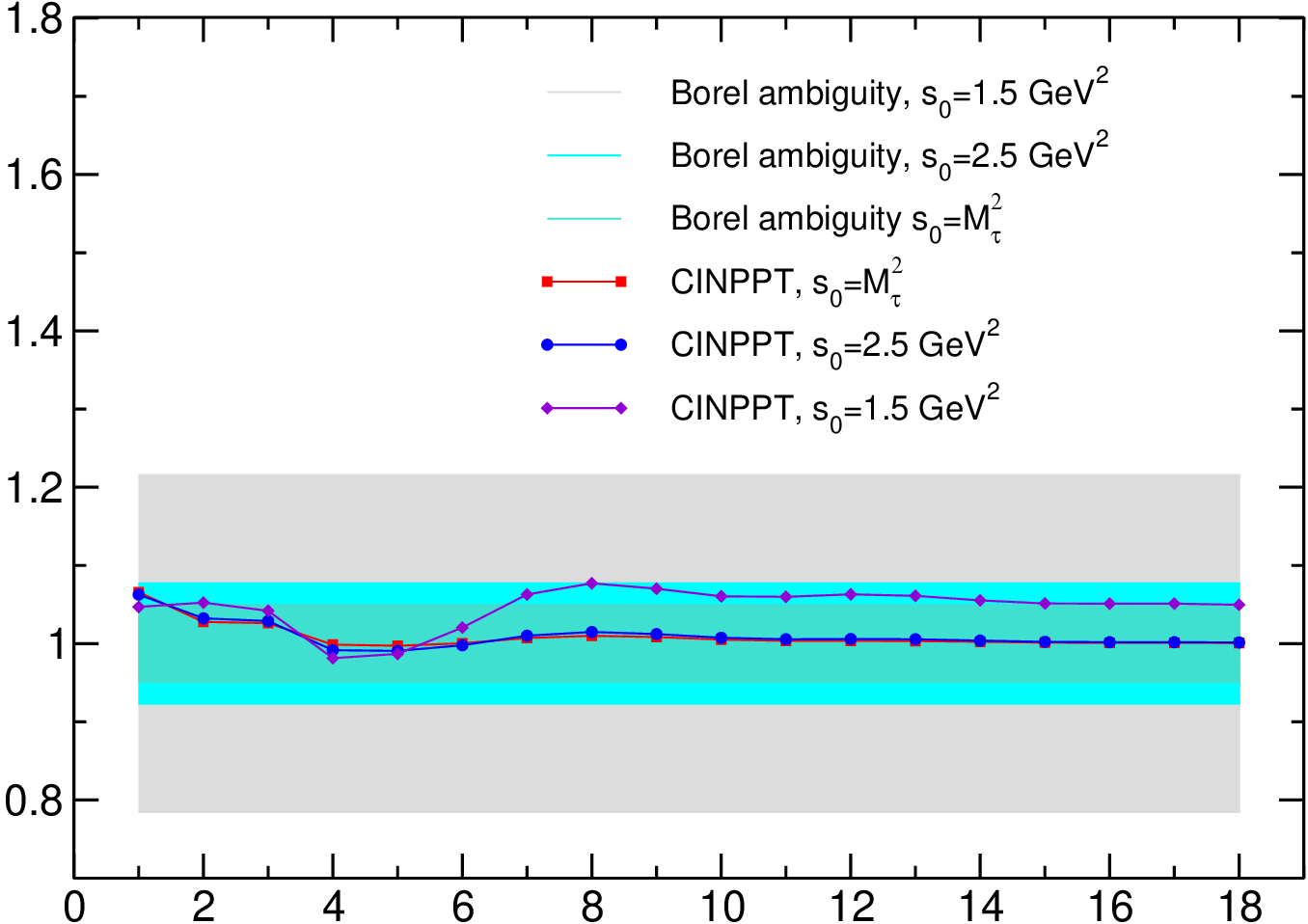}\label{w2RMs0}}\\

\vspace{.5cm}
\subfigure[~$W_{6}$, reference model]{\includegraphics[width=.8\columnwidth,angle=0]{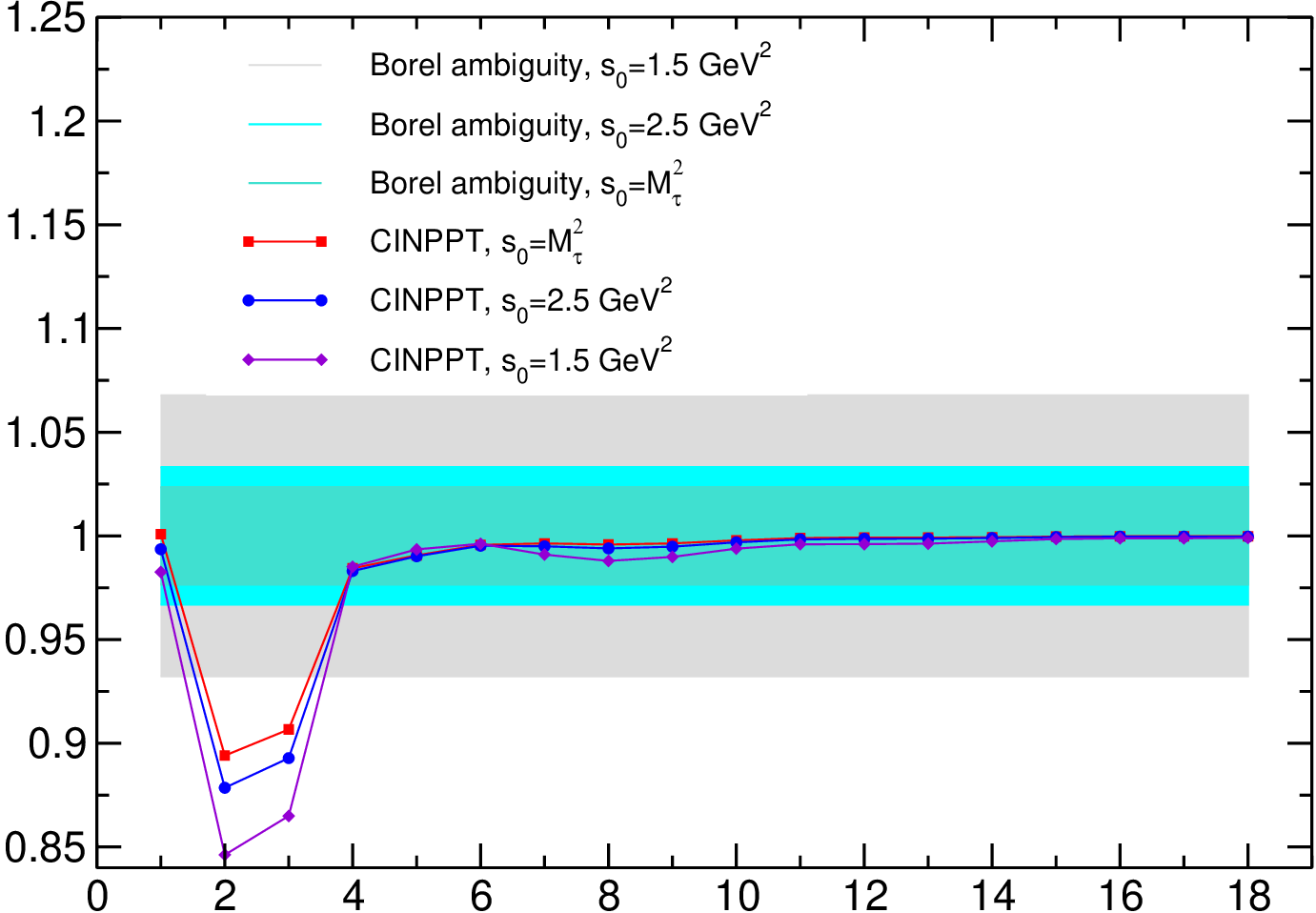}\label{w6RMs0}}\hspace{1cm}
\subfigure[~$W_{12}$, reference model]{\includegraphics[width=.8\columnwidth,angle=0]{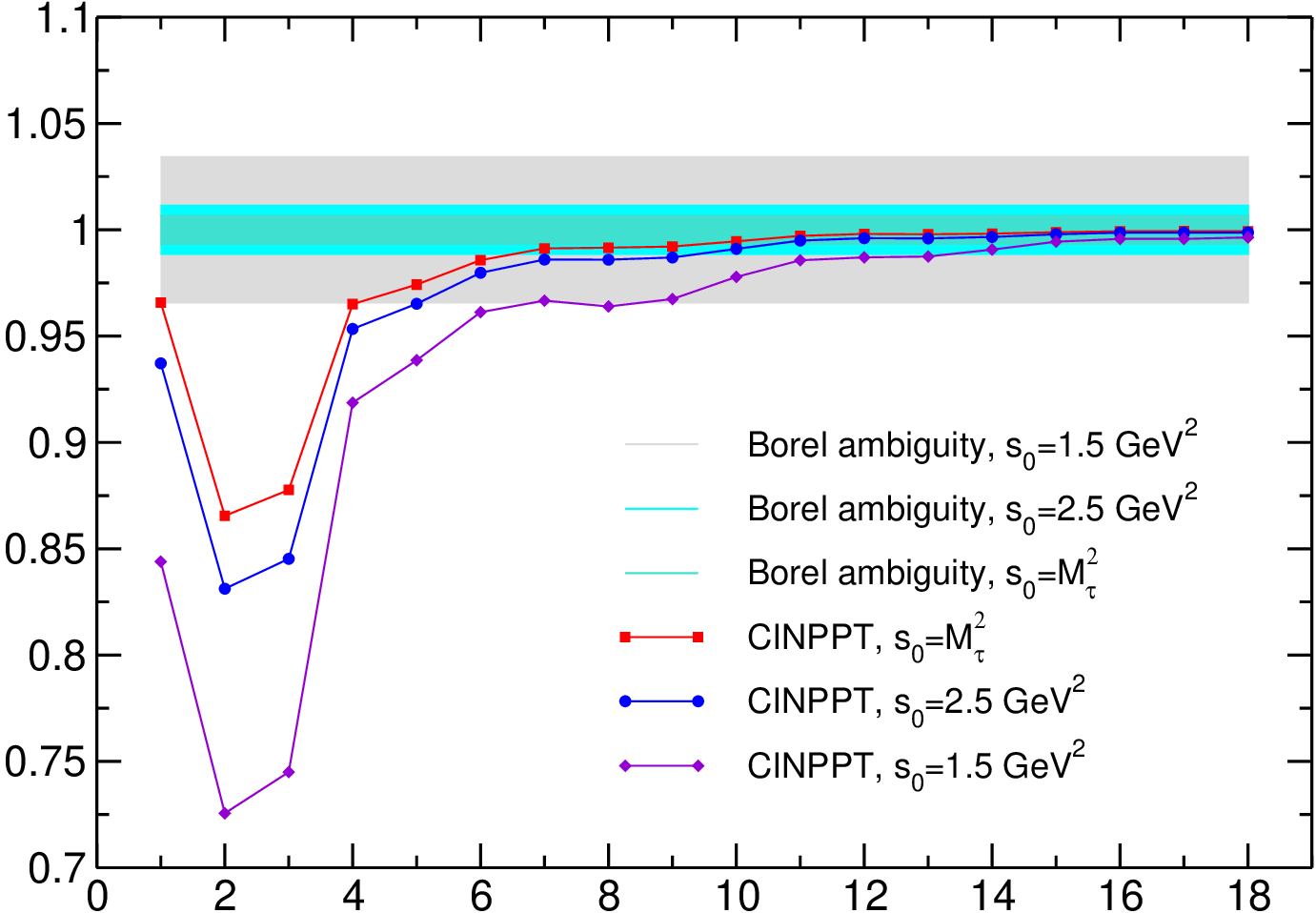}\label{w12RMs0}}\\

\vspace{.6cm}
\subfigure[~$W_{13}$, reference model]{\includegraphics[width=.8\columnwidth,angle=0]{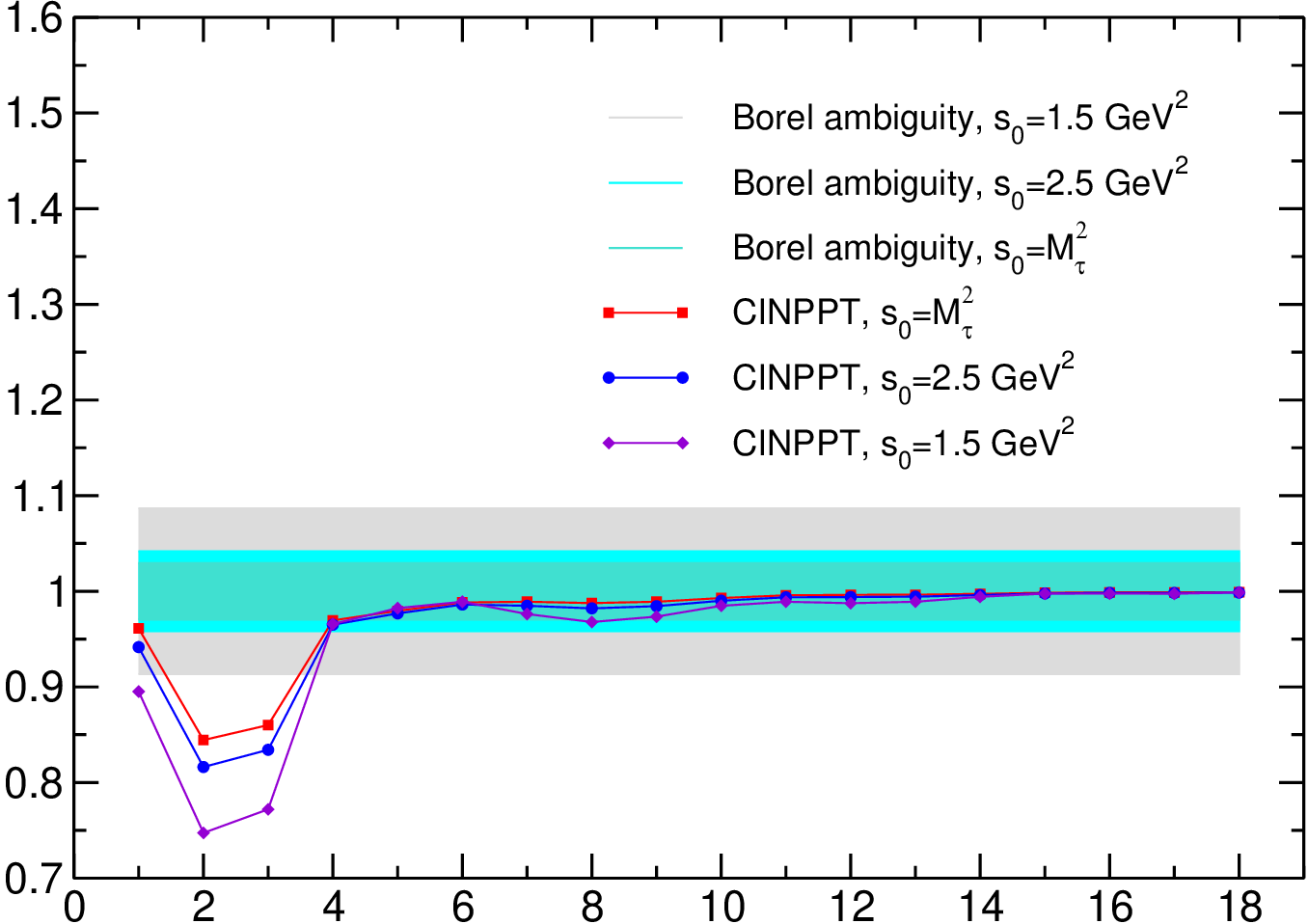}\label{w13RMs0}}\hspace{1cm}
\subfigure[~$W_{16}$, reference model]{\includegraphics[width=.8\columnwidth,angle=0]{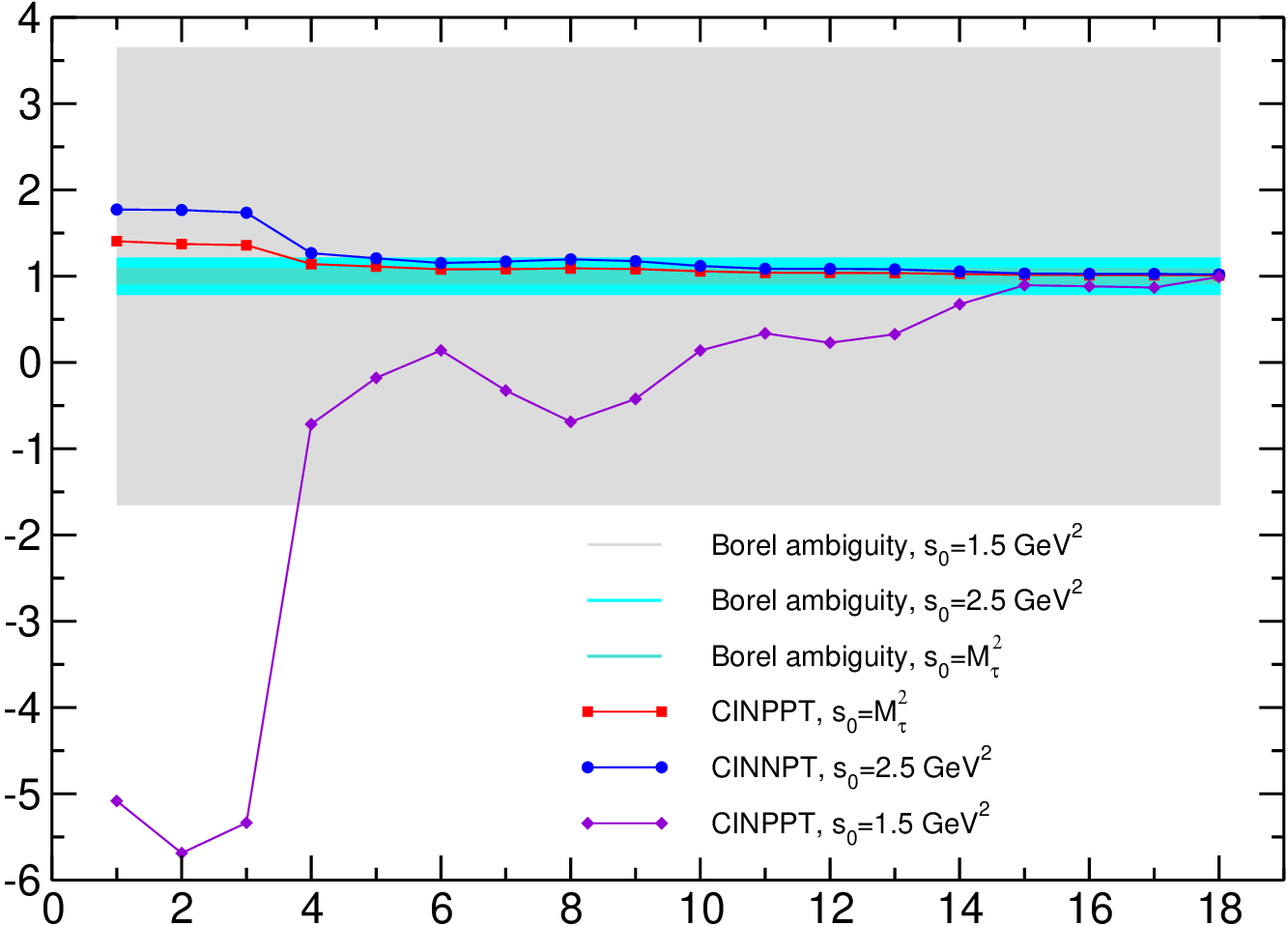}\label{w116RMs0}}
\caption{$\delta^{(0)}_{w_i}$ defined in (\ref{eq:del0}) for the RM, for $s_0=1.5\,\gev^2, 2.5\,\gev^2$ and $M_\tau^2$, expanded in the optimal CINPPT normalized to the exact value.  The horizontal bands show the uncertainties of the exact values.   }\label{fig:4}
\end{center}\end{figure*}

In the present analysis we  consider, in addition to the 
standard FOPT and CIPT series, the standard RGSPT and the 
nonpower expansions, CINPPT and RGSNPPT, defined in Sec.\ref{sec:nppt}. 
For illustration we selected several moments representative 
for each family of weights listed in Table \ref{tab:Wi}. 
The expressions given in the previous section for RM and AM were used to
compute the exact values of the moments and their theoretical uncertainties,
obtained from the imaginary part of the Laplace-Borel integral (\ref{eq:pv}) using  Eq. (A.8) of 
\cite{BeJa}. 
 The results are presented 
in Figs. \ref{fig:1},\ref{fig:4}, where we compare the perturbative expansions with the exact values and 
 their theoretical uncertainties, represented as bands. 
To facilitate the comparison with Refs. \cite{BBJ,Boito2013}, we have
 used in the calculations 
$\alpha_s(M_\tau^2)=0.3186$. 
\subsection{Results obtained with the OCM for $s_0=M_\tau^2$ }

\begin{figure*}\begin{center}

\vspace{.5cm}
\subfigure[~$W_1$, reference model]{\includegraphics[width=.8\columnwidth,angle=0]{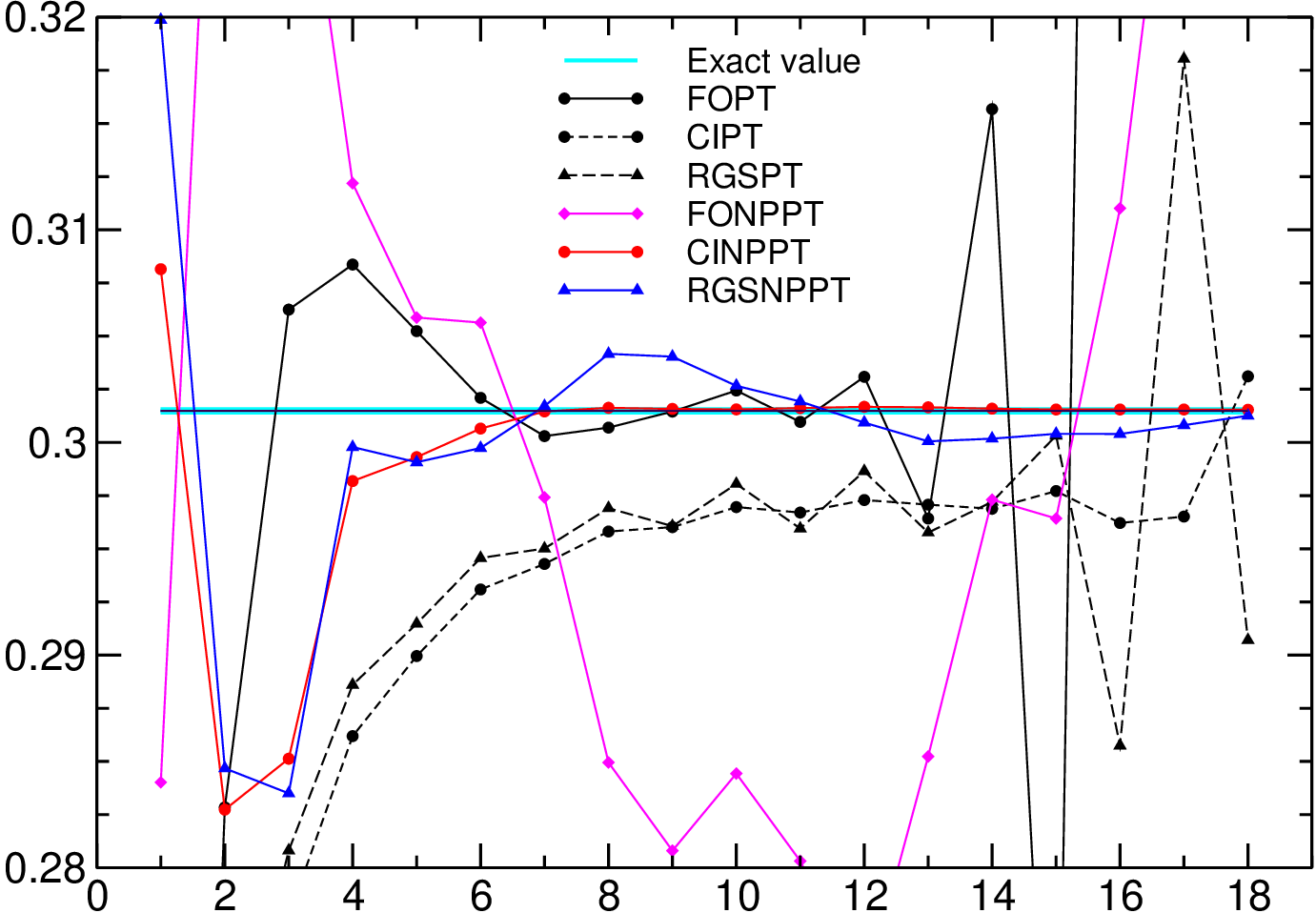}\label{w1RM}}\hspace{1cm}
\subfigure[~$W_2$, reference model]{\includegraphics[width=.8\columnwidth,angle=0]{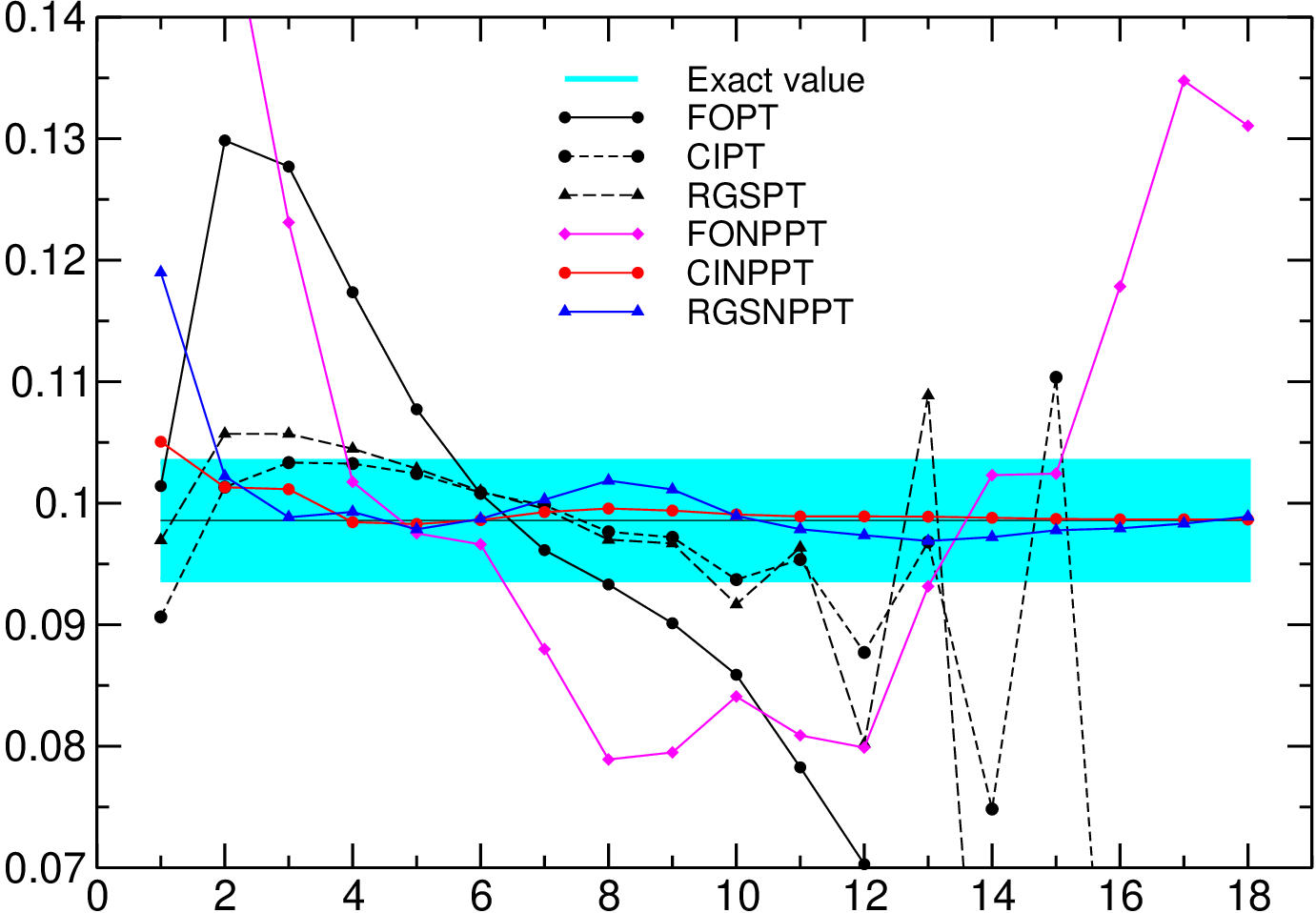}\label{w2RM}}\\

\vspace{.5cm}
\subfigure[~$W_{6}$, reference model]{\includegraphics[width=.8\columnwidth,angle=0]{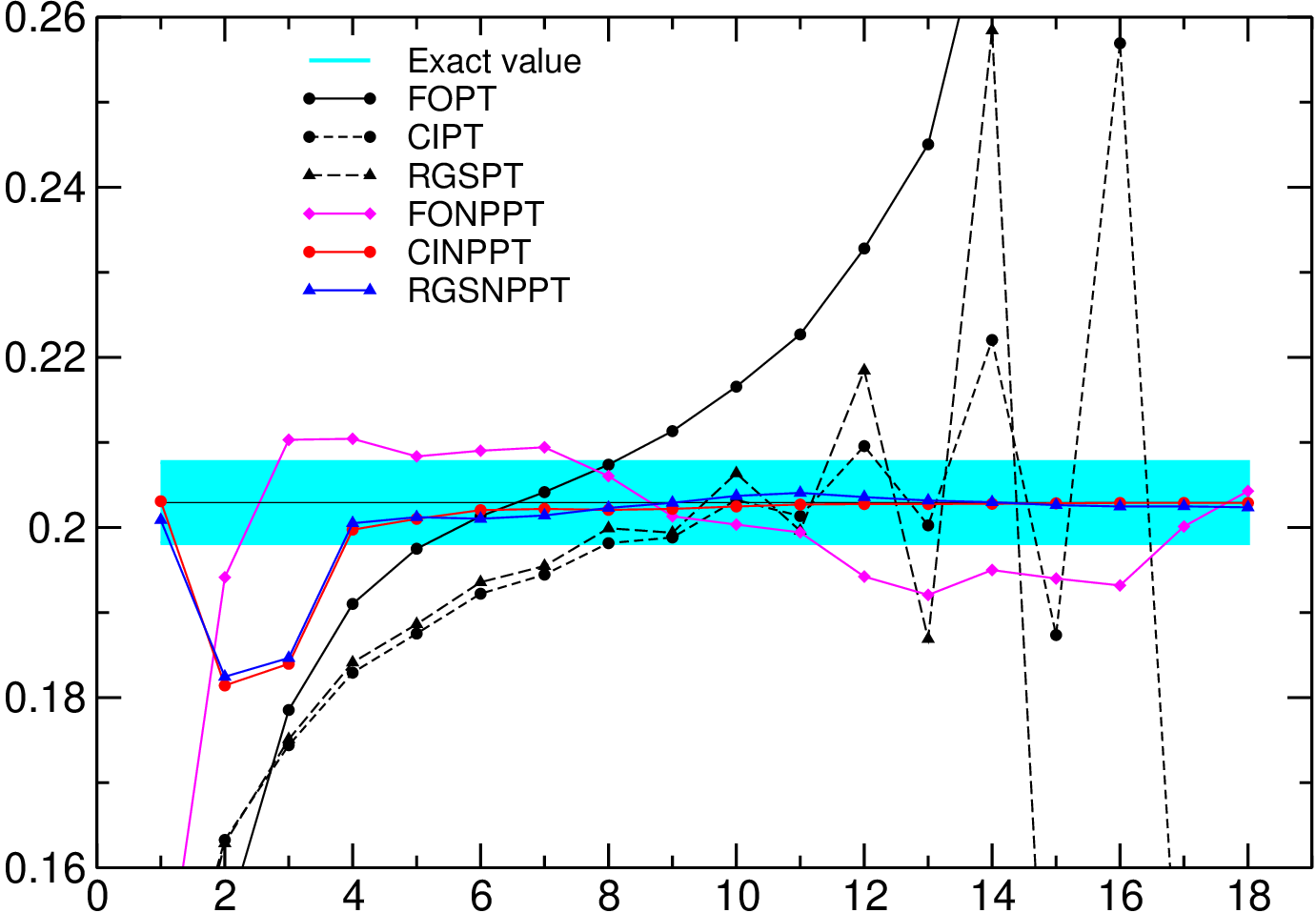}\label{w6RM}}\hspace{1cm}
\subfigure[~$W_{12}$, reference model] {\includegraphics[width=.8\columnwidth,angle=0]{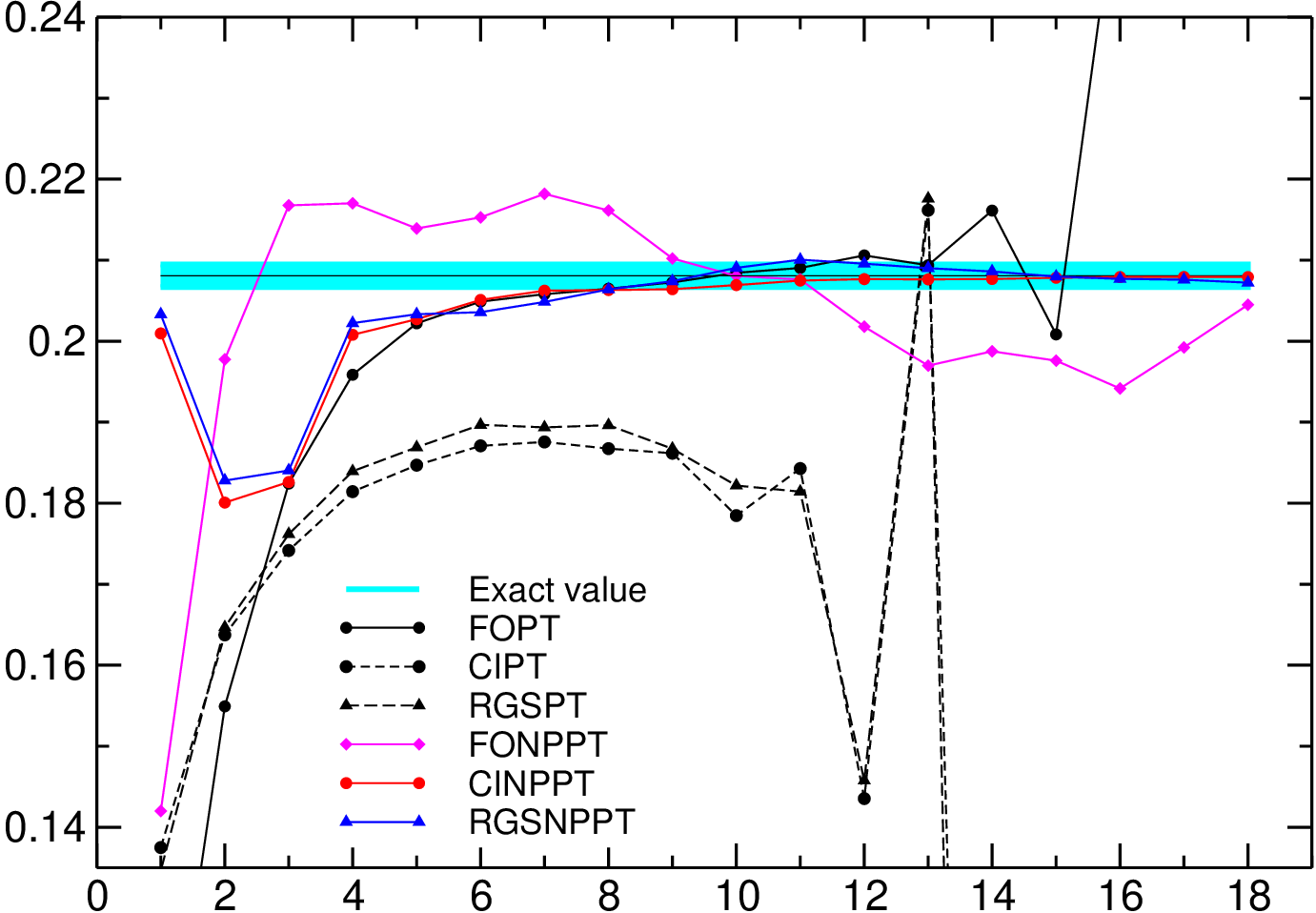}\label{w12RM}}\\

\vspace{.5cm}
\subfigure[~$W_{13}$, reference model]{\includegraphics[width=.8\columnwidth,angle=0]{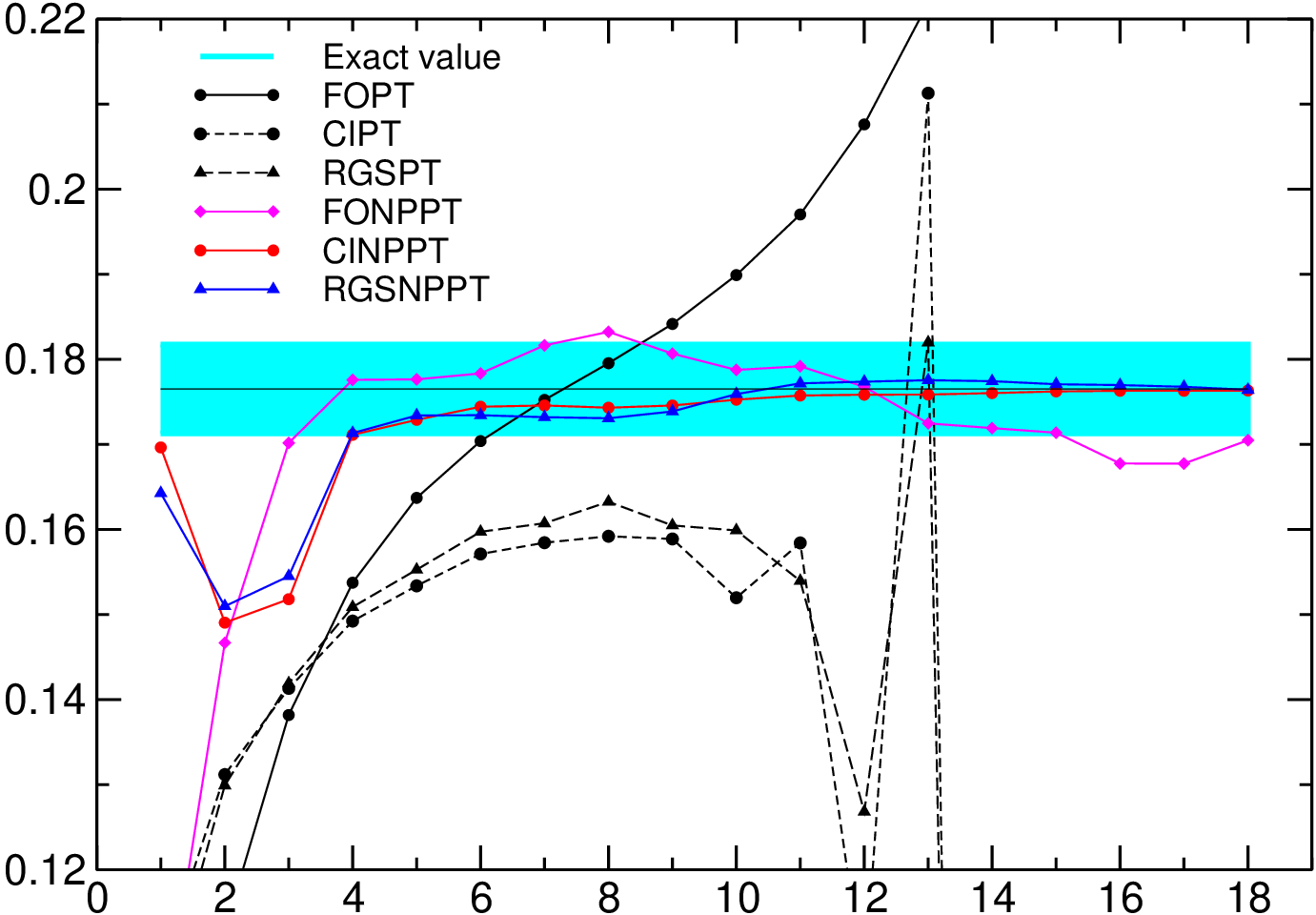}\label{w13RM}}\hspace{1cm}
\subfigure[~$W_{16}$, reference model]{\includegraphics[width=.8\columnwidth,angle=0]{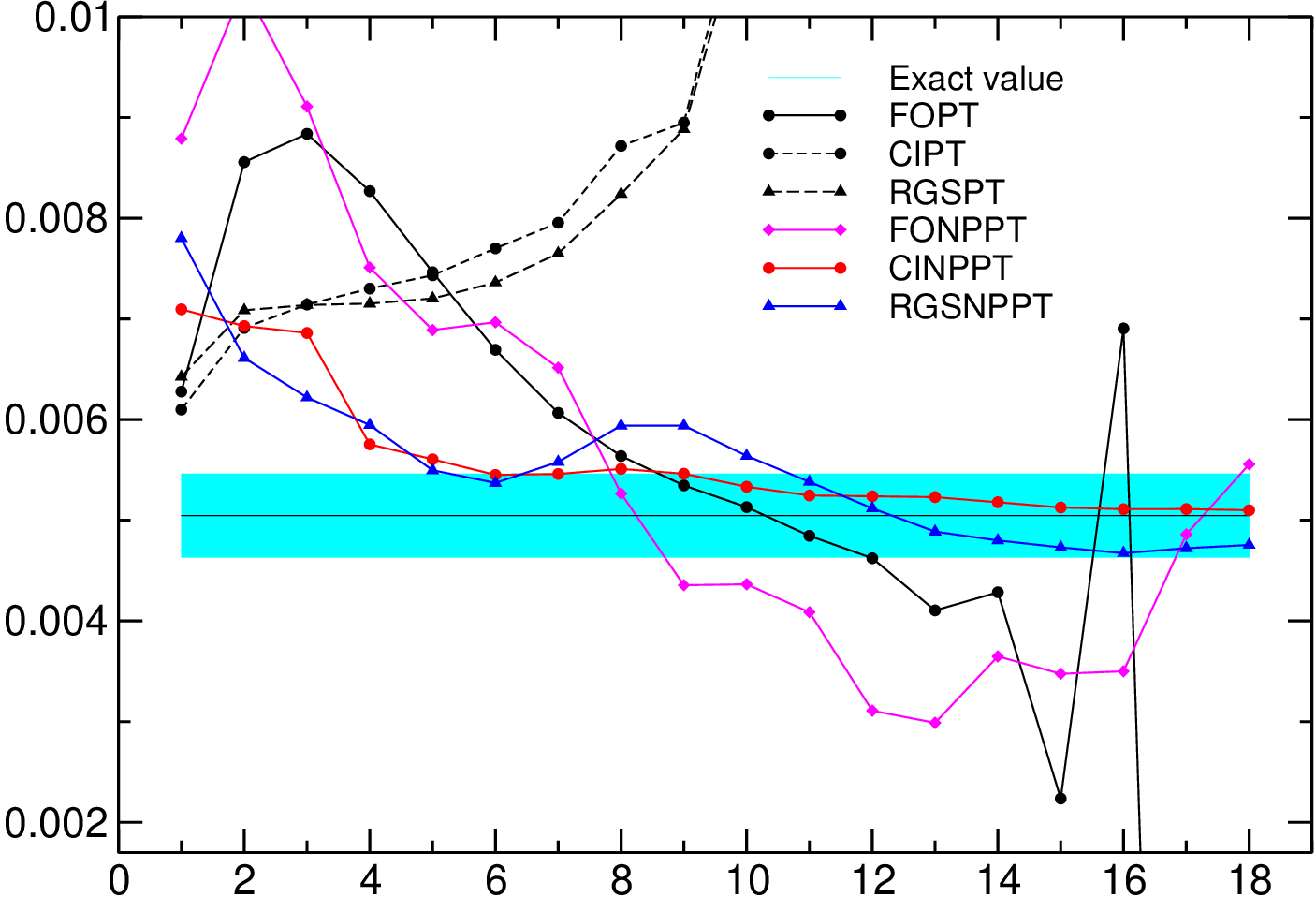}\label{w16RM}}
\caption{$\delta^{(0)}_{w_i}$ defined in (\ref{eq:del0}) for the weights $W_1$, $W_2$, $W_6$, $W_{12}$, $W_{13}$ and $W_{16}$, calculated for the RM 
defined in \cite{BeJa,BBJ} with the standard and  nonpower versions of FO, CI and RGS expansions, as functions of the perturbative order up to which the series was summed. The horizontal 
bands give the uncertainties of the exact values. As in \cite{BBJ}, we use $\alpha_s(M_\tau^2)=0.3186$.}\label{fig:1}
\end{center}\end{figure*}

We investigate first  the nonpower expansions 
obtained from (\ref{eq:cinppt}) and (\ref{eq:rgsnppt}) with the choice $j=1$ 
and $k=2$, which  corresponds to the OCM 
defined in Sec. \ref{sec:nppt}. For the softening factor $S(u)$ we adopt the expression given in (\ref{eq:Sjk}) for the same values of $j$ and $k$.
In  Fig. \ref{fig:1}  we show the perturbative behavior of the 
chosen sample of moments in the case of RM.  For each moment we show the behavior of the standard expansions FOPT, CIPT and RGSPT, together with the optimal nonpower 
version of each expansion.

As  noted already from studies of the hadronic width \cite{Abbas:2012, AACSing}, 
RGSPT gives results very close to CIPT, and this is confirmed for all 
the moments shown in Fig. \ref{fig:1}.  As concerns the FO nonpower expansions, the previous studies \cite{CaFi2009,CaFi2011, CaSing} showed that they achieve a good approximation of the Adler 
function near the spacelike axis, taming the large-order increase of the leading coefficients $c_{n,1}$, but have a bad behavior near the timelike axis, where the unsummed $s$-dependent logarithms present in the coefficients are large.   Therefore, we expect good perturbative results only for "pinched" moments, like the 6th or the 12th, where the weight suppresses the region near the timelike axis. The results presented in Fig. \ref{fig:1} confirm this expectation.  Incidentally, the improvement of the large-order behavior  may destroy some suitable compensations of terms that take place in the standard FOPT and explain the good perturbative behavior of this scheme for some moments. Therefore, as already concluded in \cite{CaFi2009,CaFi2011}, the FO nonpower perturbative scheme is not suitable, because it cures only one facet of the problem, {\em i.e.} the large-order behavior, leaving the renormalization-group coefficients unsummed. To optimize the perturbative expansion, we must improve both aspects, as done within the CI and RGS nonpower frameworks.

\begin{figure*}\begin{center}

\vspace{.5cm}
\subfigure[~$W_1$, alternative model]{\includegraphics[width=.8\columnwidth,angle=0]{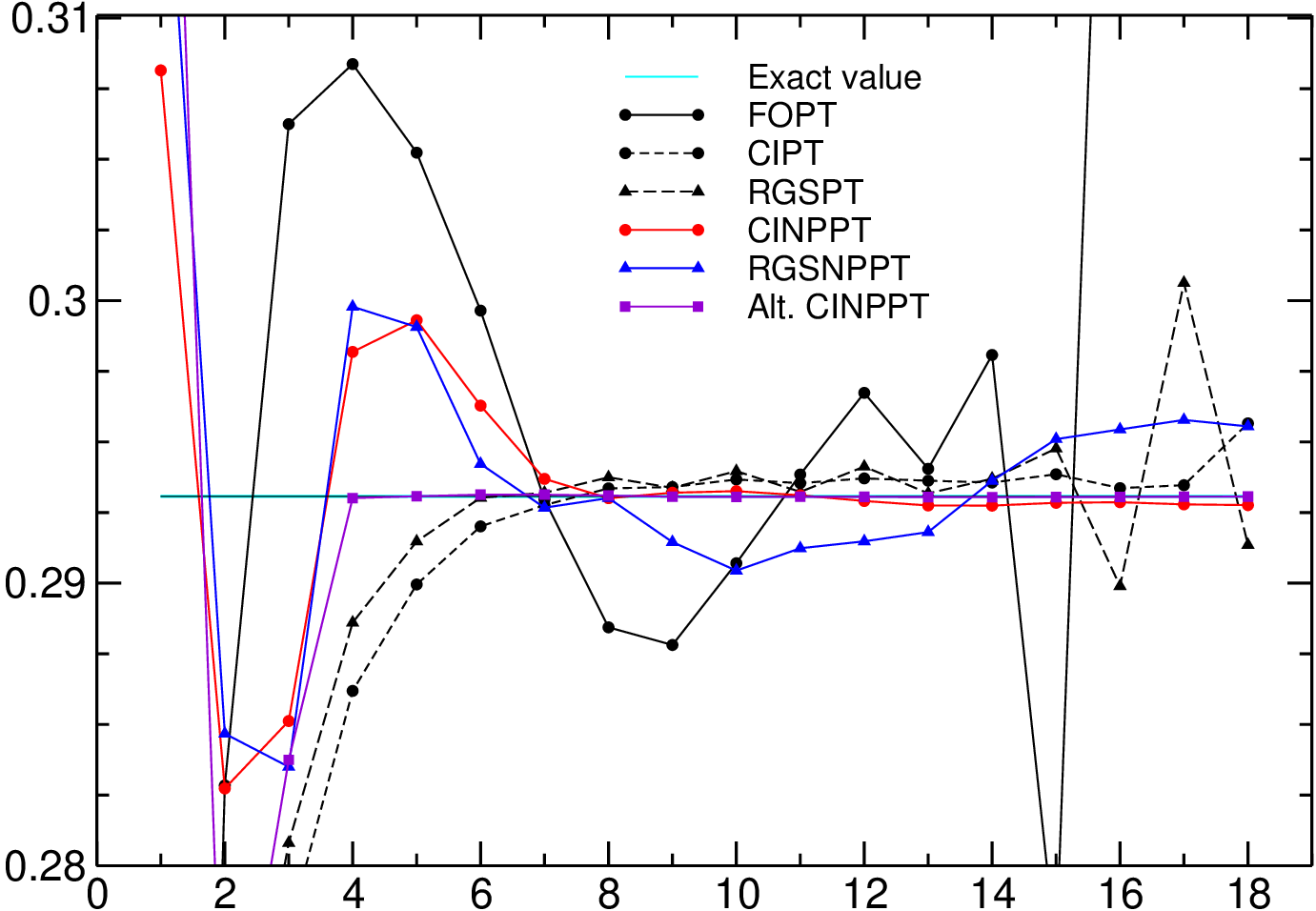}\label{w1AM}}\hspace{1cm}
\subfigure[~$W_2$, alternative model]{\includegraphics[width=.8\columnwidth,angle=0]{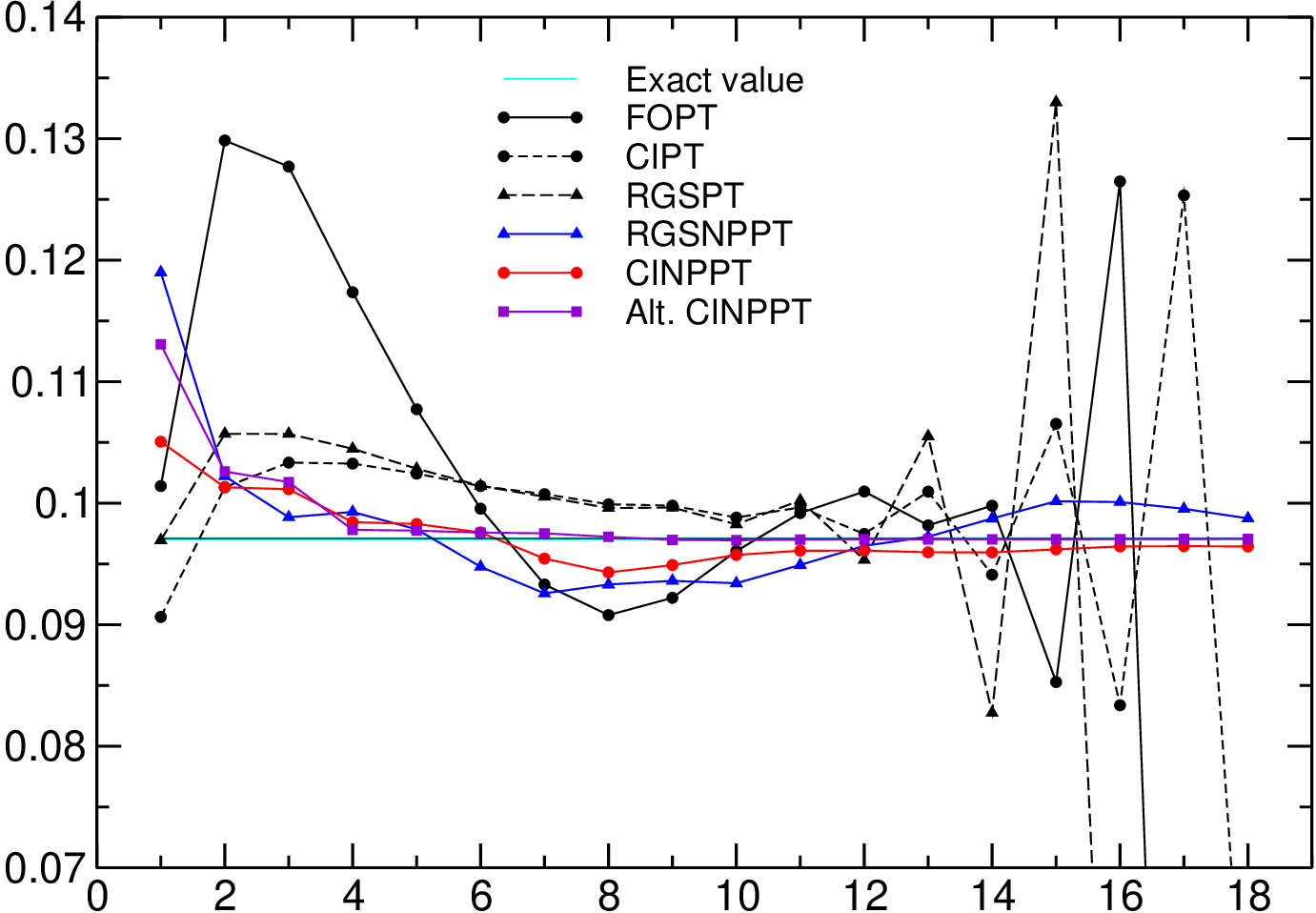}\label{w2AM}}\\

\vspace{.5cm}
\subfigure[~$W_{6}$, alternative model]{\includegraphics[width=.8\columnwidth,angle=0]{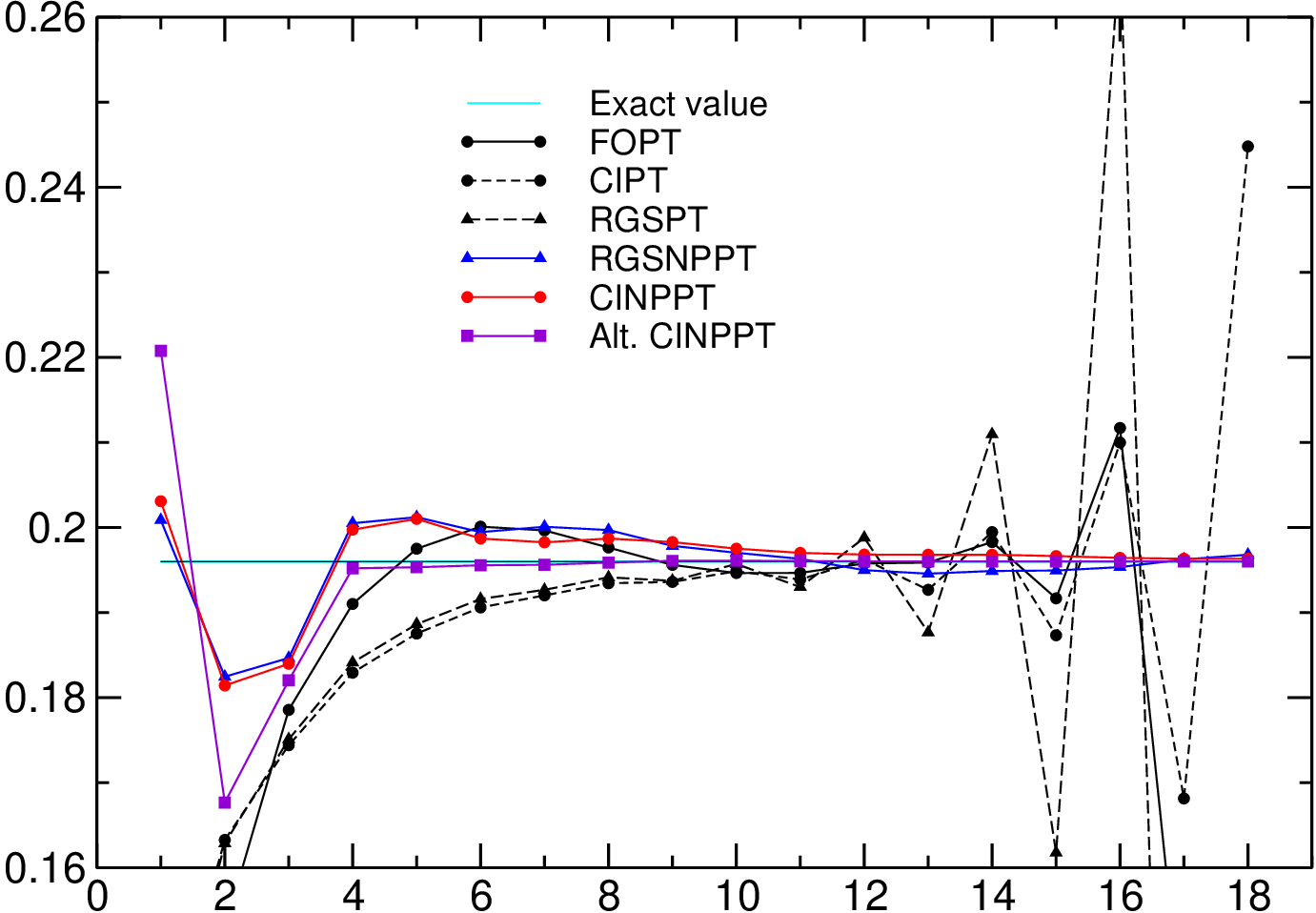}\label{w6AM}}\hspace{1cm}
\subfigure[~$W_{12}$, alternative model] {\includegraphics[width=.8\columnwidth,angle=0]{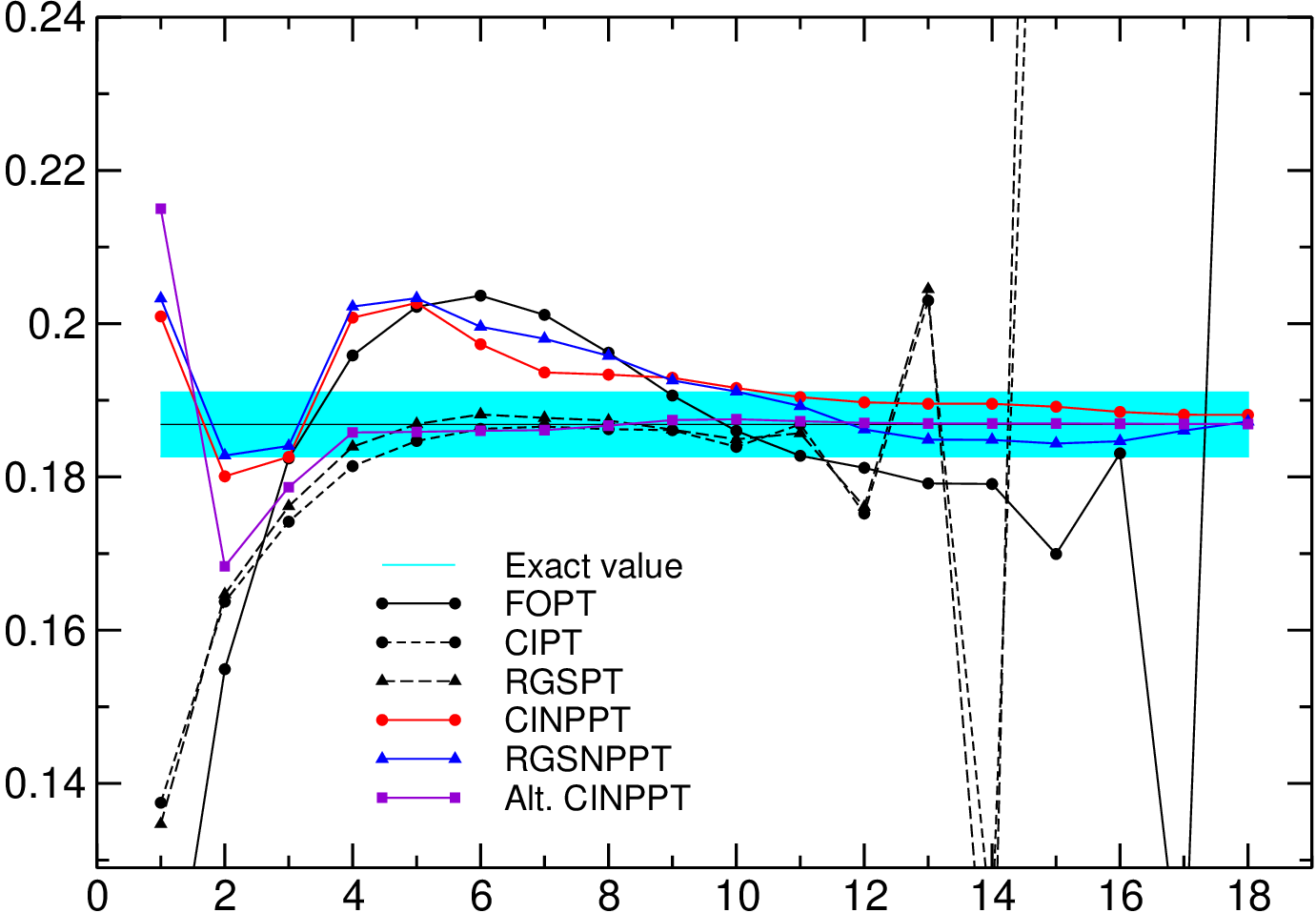}\label{w12AM}}\\

\vspace{.5cm}
\subfigure[~$W_{13}$, alternative model]{\includegraphics[width=.8\columnwidth,angle=0]{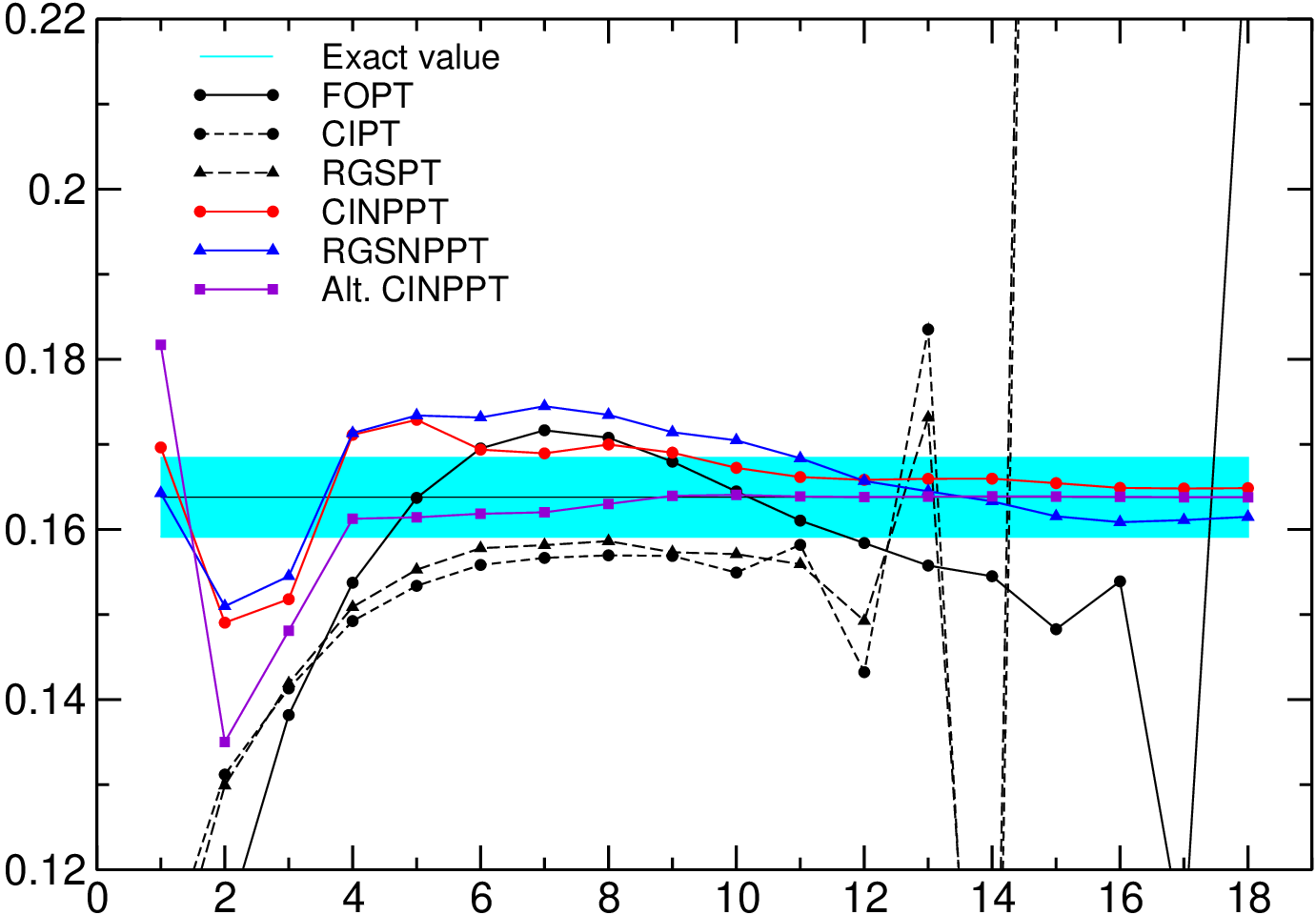}\label{w13AM}}\hspace{1cm}
\subfigure[~$W_{16}$, alternative  model]{\includegraphics[width=.8\columnwidth,angle=0]{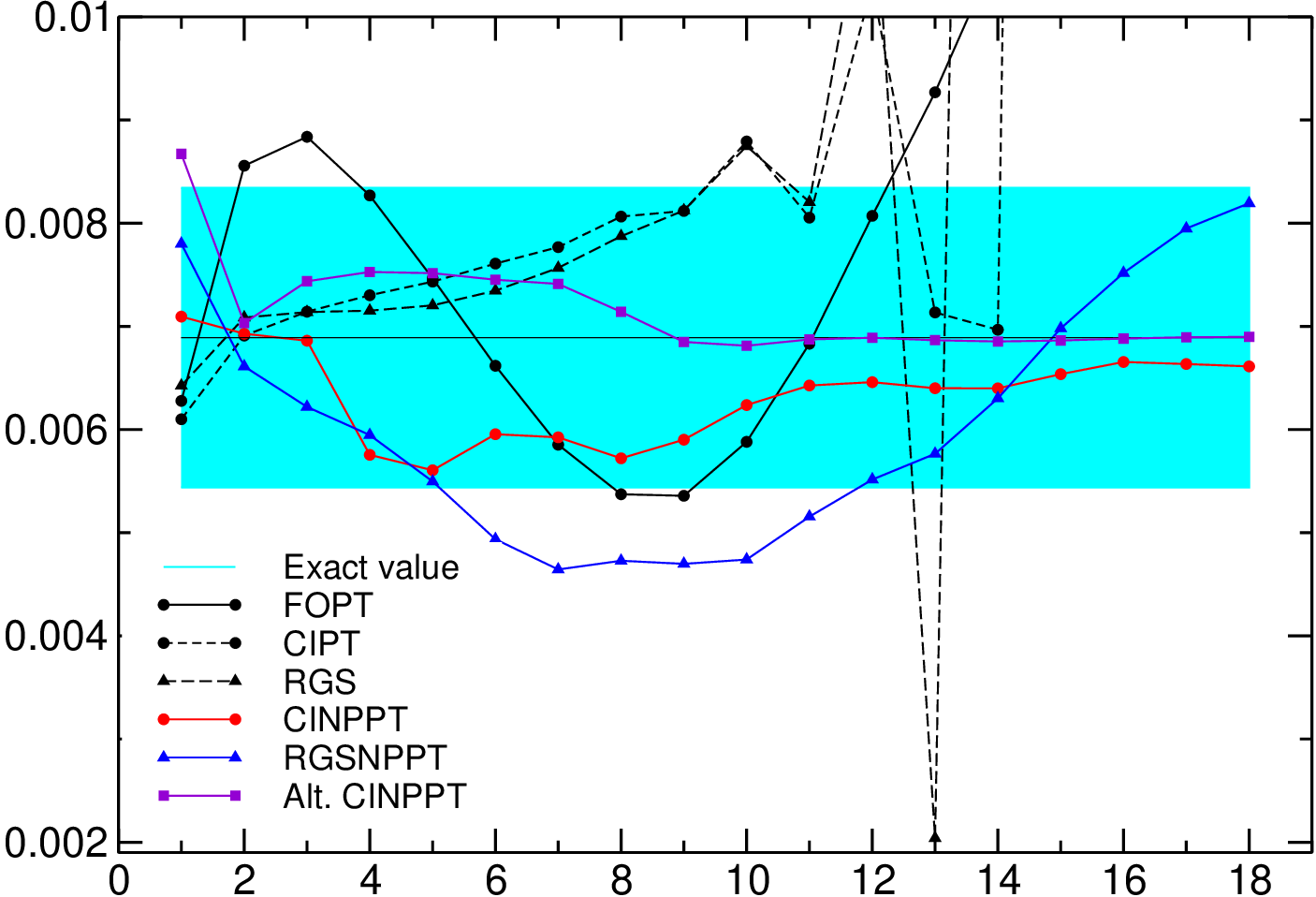}\label{w16AM}}
\caption{Perturbative expansions of the moments of the AM adopted in \cite{BBJ}.  ``Alt. CINPPT" denotes the specific optimal expansion devised for the AM, as explained in the text. }\label{fig:2}
\end{center}\end{figure*}

The remaining two curves in each figure, denoted as CINPPT and RGSNPPT,  
prove in an impressive way the excellent approximation achieved 
with the CI and RGS nonpower expansions based on the OCM 
$\wt w_{12}(u)$, even for moments for which the standard CI, FO and RGS 
expansions fail badly. The only moment for which the  perturbative description is less
 impressive at low and moderate orders is the one obtained with 
the weight $W_{16}$. However, this moment is very small and has a 
large uncertainty, so the description 
may be considered good in this case as well.   In all the cases, one may 
note a slightly better description achieved by CINPPT compared to RGSNPPT. 
The other moments defined in Table \ref{tab:Wi}, for which we do not
explicitly exhibit the results, have a  behavior similar to that 
of the representative moment of their class.

 The good convergence of CINPPT and RGSNPPT for the moments shown in Fig. \ref{fig:1}
can be understood from previous studies  \cite{CaFi2011,AACF,CaSing}, 
which demonstrated that these expansions provide a very good approximation 
of the exact Adler function itself in the complex plane along the whole 
circle $|s|= M_\tau^2$. The good pointwise convergence of these expansions implies a good convergence to the true values
 also for contour integrals defined in (\ref{eq:del0}), for all types of weights $W_i(x)$. 

We consider now the AM discussed in \cite{BBJ}, 
specified above in Eqs. (\ref{eq:altBBJ}) and (\ref{eq:altdBJ}). 
We recall that in this extreme model the first singularity of the 
Borel transform at $u=2$ is completely removed. On the other hand, 
the nonpower  expansions defined in Sec. \ref{sec:nppt}  explicitly
implement both the position and the nature of this singularity, 
known theoretically. In particular, the expansion functions (\ref{eq:Wnpci}) explicitly
contain the singularity at $u=2$ in the Borel plane
(known actually to be present in the true, physical Adler function), 
while the function that we want to approximate  does not have 
such a singularity.  This means that the expansion functions defined 
in Sec. \ref{sec:nppt} are not mathematically optimal  for this extreme model.  
We expect therefore a slower convergence  
and a poorer description of the true values at low orders.

On the other hand, after the conformal mapping of the 
cut $u$ plane onto the unit disk, the  expansion (\ref{eq:Bw}) 
of the Borel function converges in a larger domain. This leads 
also to a better convergence at large-orders for points $u$ on the real axis
near the origin, which dominate the Laplace-Borel integral. 
Therefore, we expect the nonpower expansions defined 
in Sec. \ref{sec:nppt} to exhibit a tame behavior 
at large-orders also in the case of the AM. 

These expectations are confirmed by the
results shown in Fig. \ref{fig:2}, where 
we present the moments considered in Fig. \ref{fig:1} 
for the AM: at high orders  
the nonpower expansions  tend to  the exact value, 
illustrating the  series acceleration by the 
OCM \cite{CiFi, CaFi2011}. 
The description is relatively good even at low orders for weights like $W_2$ and $W_6$, for
 which the exact values of the moments in RM and AM are rather close (however the uncertainty of these moments in AM is much smaller, requiring a better precision). For other moments, for which the true values in AM are quite different from  those in RM, the approximation at low orders is worse in AM compared to RM.

In order to gain further insight, we have also carried out
a study of the CINPPT series for the Adler function in the complex plane, along the contour $|s|=s_0$. Note that the true Adler function defined by the AM has a more oscillating behavior along the circle compared to
the RM (this was noted also for other models in Ref. \cite{BBJ}).  For large perturbative orders $N$, the  series approaches
the true values for both the real and imaginary parts of $\wh D(s)$ quite uniformly along the circle. On the other hand, at low orders, 
the expansions (which are the same for all models up to $N= 5$), stay quite close to the true function defined by the RM, 
departing therefore  from the AM. In particular, they are not able to reproduce the oscillations of the model along the circle. This 
shows that  the CINPPT expansions approximate better the exact Adler function defined by the RM than the function defined by the AM.

As we mentioned above, the expansion that we used is not optimal for the AM. One can actually define an optimal expansion for this model, using the fact that
its first singularities are situated at $u=-1$ and $u=3$, 
and have a known nature \cite{BeJa,BBJ}. The  optimal mapping is obtained by setting
 $j=1$ and $k=3$ in (\ref{eq:wjk}). Moreover,
the softening factor $S(u)$ must vanish at $u=-1$ and $u=3$. 
Adopting for $S(u)$ the expression (\ref{eq:Sjk}), we obtain the proper factor by replacing $\wt w_{12}(2)$ and 
$\gamma'_2$ by  
$\wt w_{13}(3)$ and the  value of $\gamma'_3$ derived from the 
parameter $\gamma_3$ given in  \cite{BeJa, BBJ}.  
It is instructive to show also the results obtained with this  optimal perturbative
expansion suitable for the AM. It is denoted as ``Alt. CINPPT" 
in Fig. \ref{fig:2}, and exhibits a very rapid convergence 
to the true values of all the moments. 

This exercise demonstrates the mathematical 
power of the technique of the singularity softening and  
conformal mappings for series acceleration when the position 
and the nature of the leading singularities is known. 
Of course, for the physical Adler function, where the first IR renormalon is known to be present, the softening factor $S(u)$ must vanish at the dominant branch points  $u=-1$ and $u = 2$, and the optimal expansion variable must map 
onto a disk the $u$ plane cut along $u\leq -1$ and $u \ge 2$.

\begin{figure*}\begin{center}

\vspace{.5cm}
\subfigure[~$W_1$, reference model]{\includegraphics[width=.8\columnwidth,angle=0]{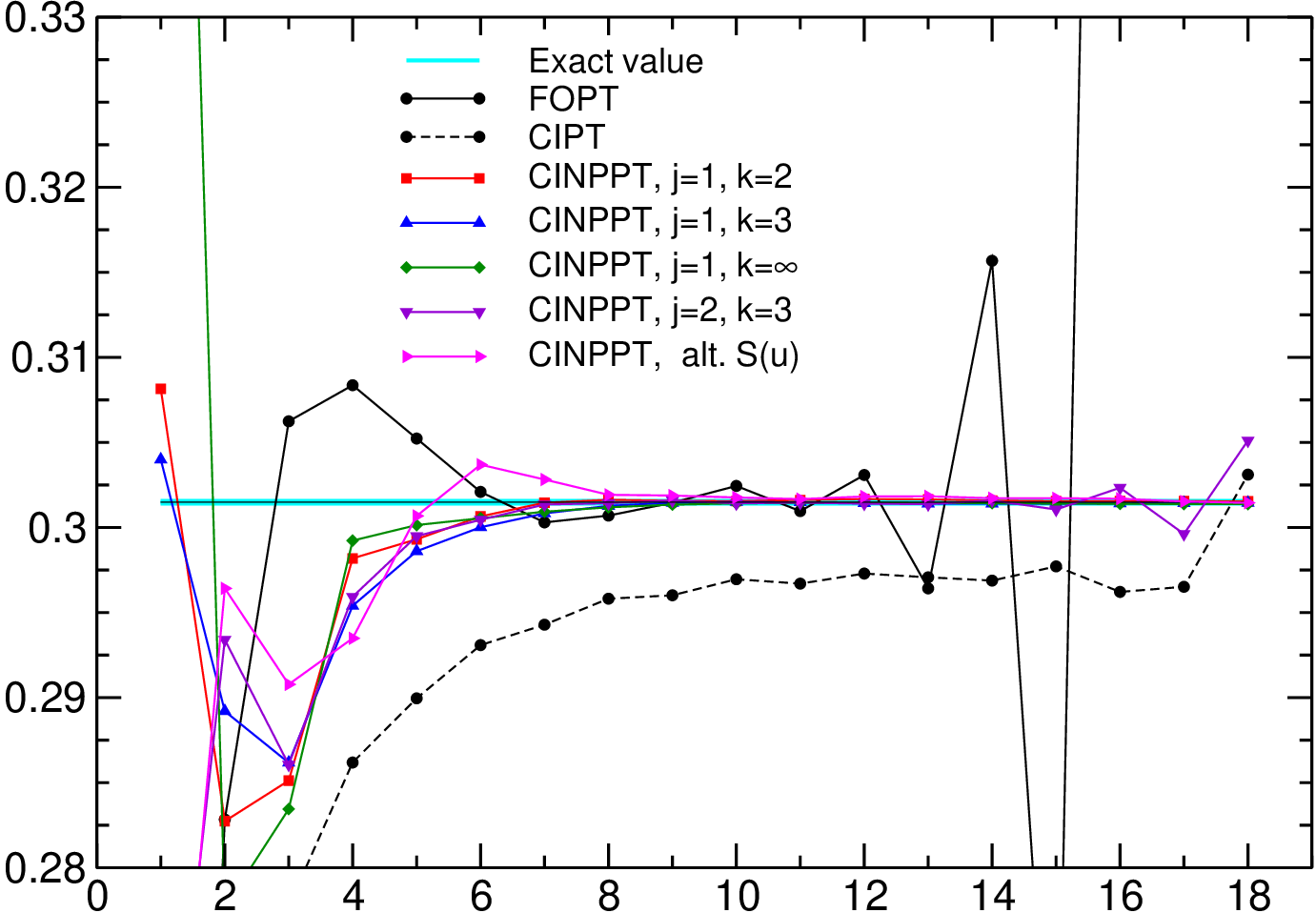}\label{w1RMw}}\hspace{1cm}
\subfigure[~$W_2$, reference model]{\includegraphics[width=.8\columnwidth,angle=0]{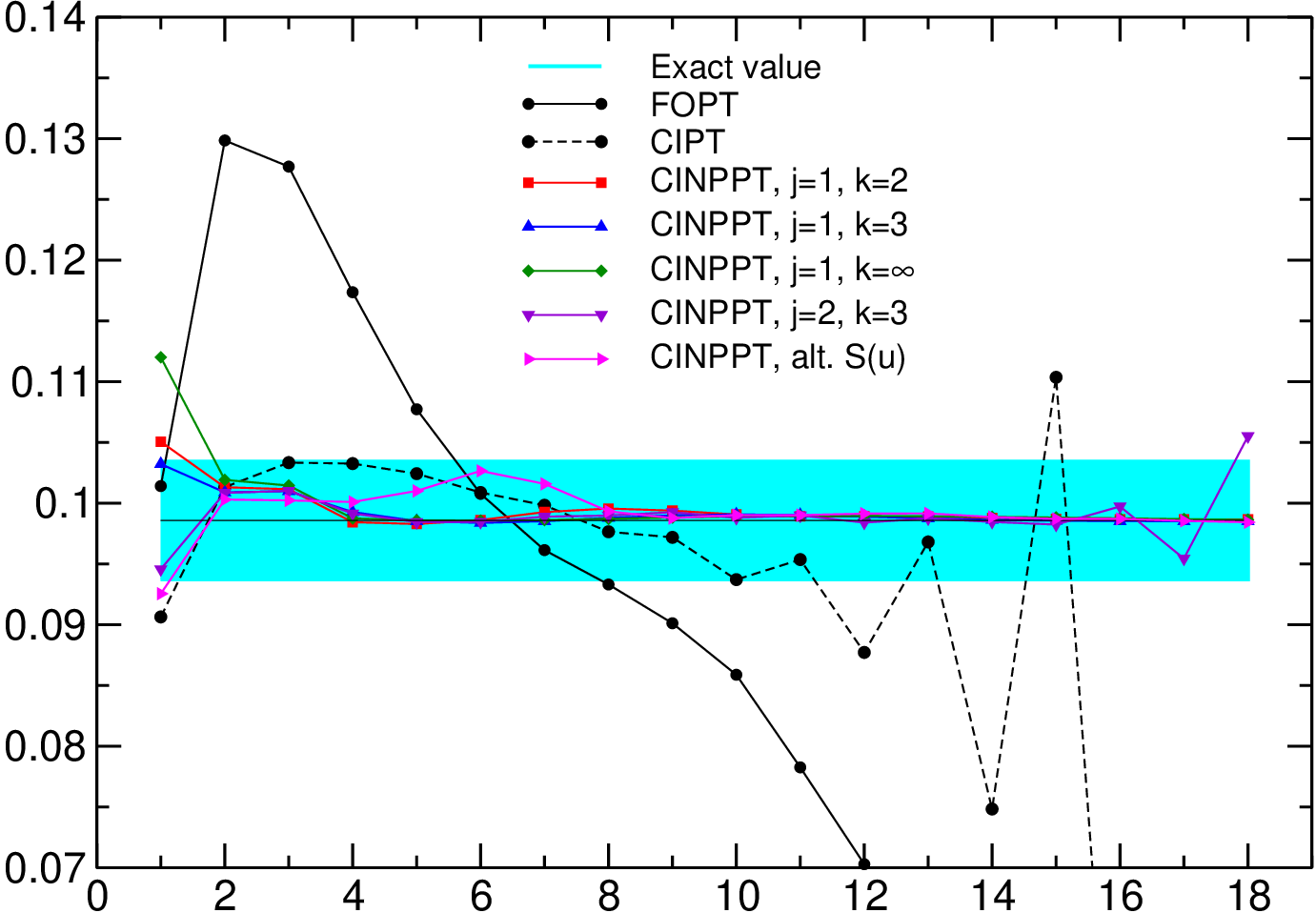}\label{w2RMw}}\\

\vspace{.5cm}
\subfigure[~$W_{6}$, reference model]{\includegraphics[width=.8\columnwidth,angle=0]{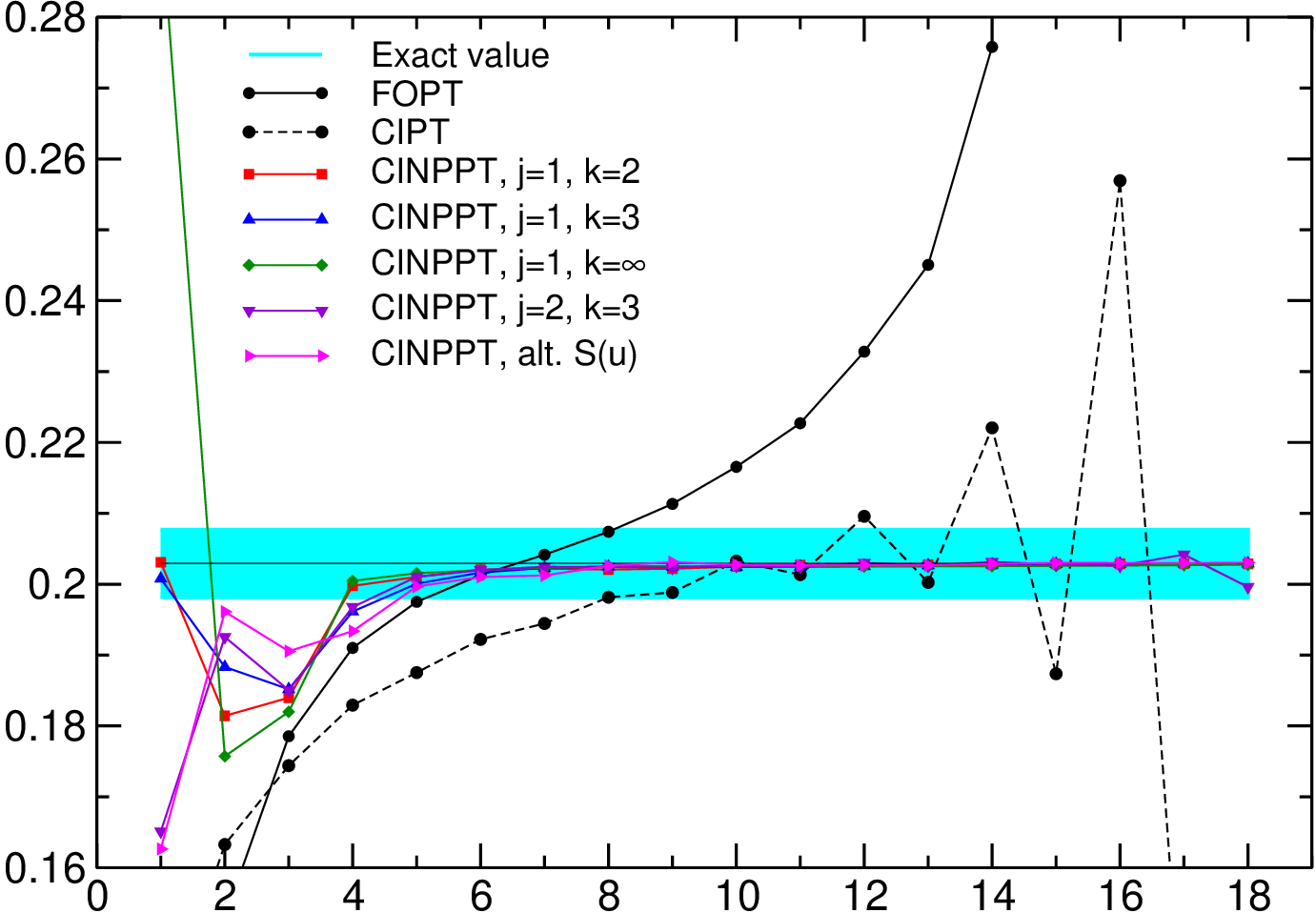}\label{w6RMw}}\hspace{1cm}
\subfigure[~$W_{12}$, reference model] {\includegraphics[width=.8\columnwidth,angle=0]{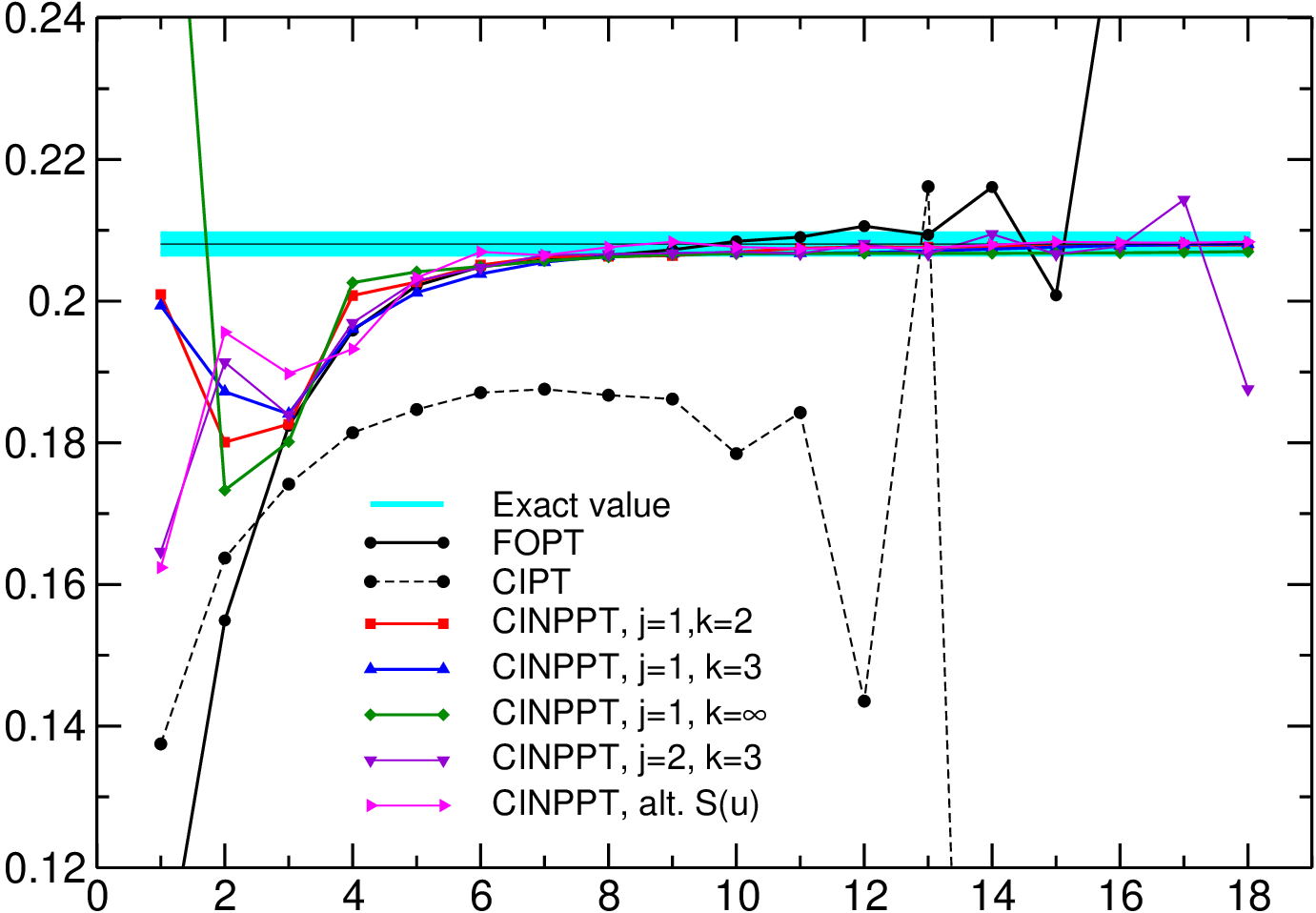}\label{w12RMw}}\\

\vspace{.5cm}
\subfigure[~$W_{13}$, reference model]{\includegraphics[width=.8\columnwidth,angle=0]{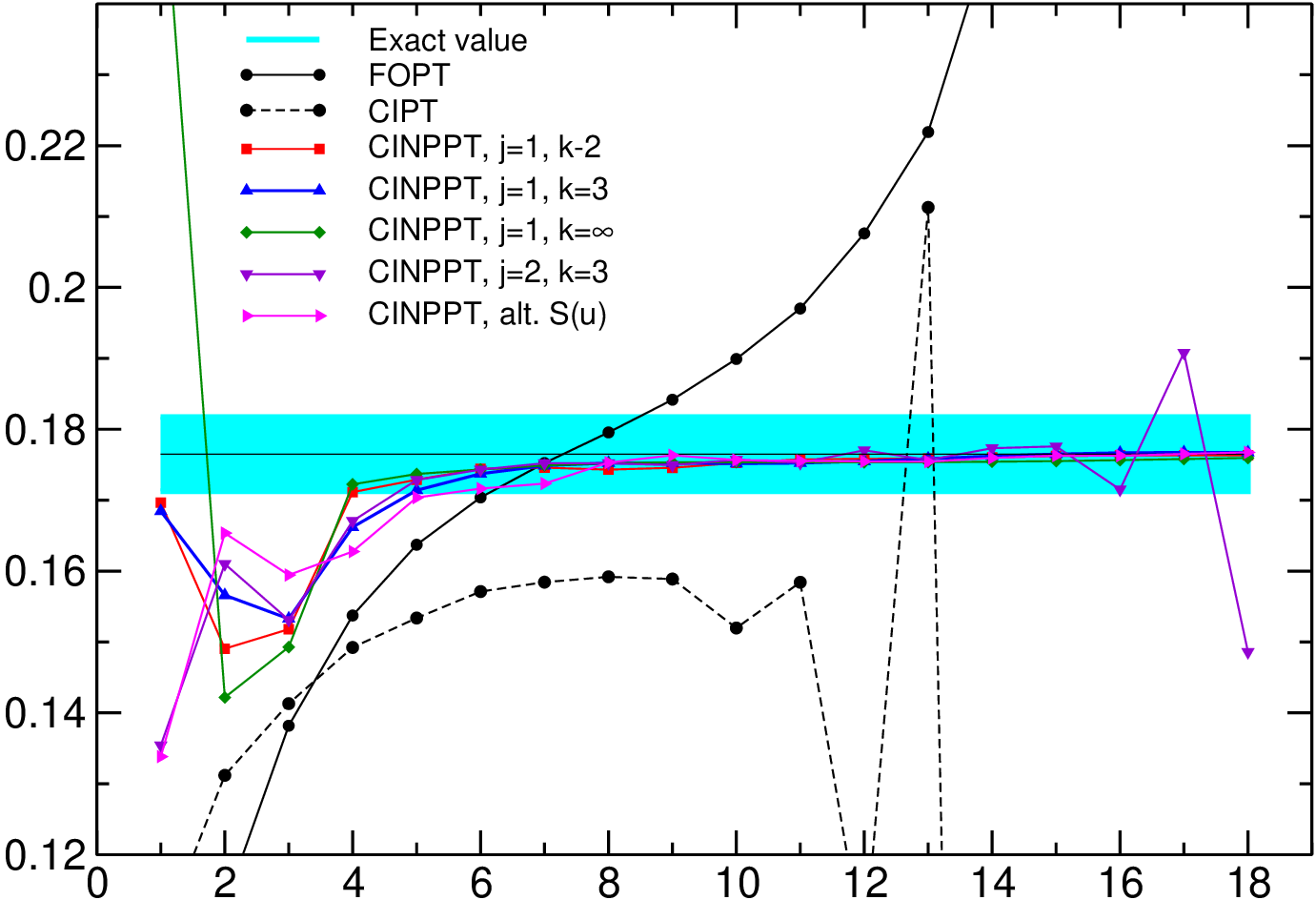}\label{w13RMw}}\hspace{1cm}
\subfigure[~$W_{16}$, reference model]{\includegraphics[width=.8\columnwidth,angle=0]{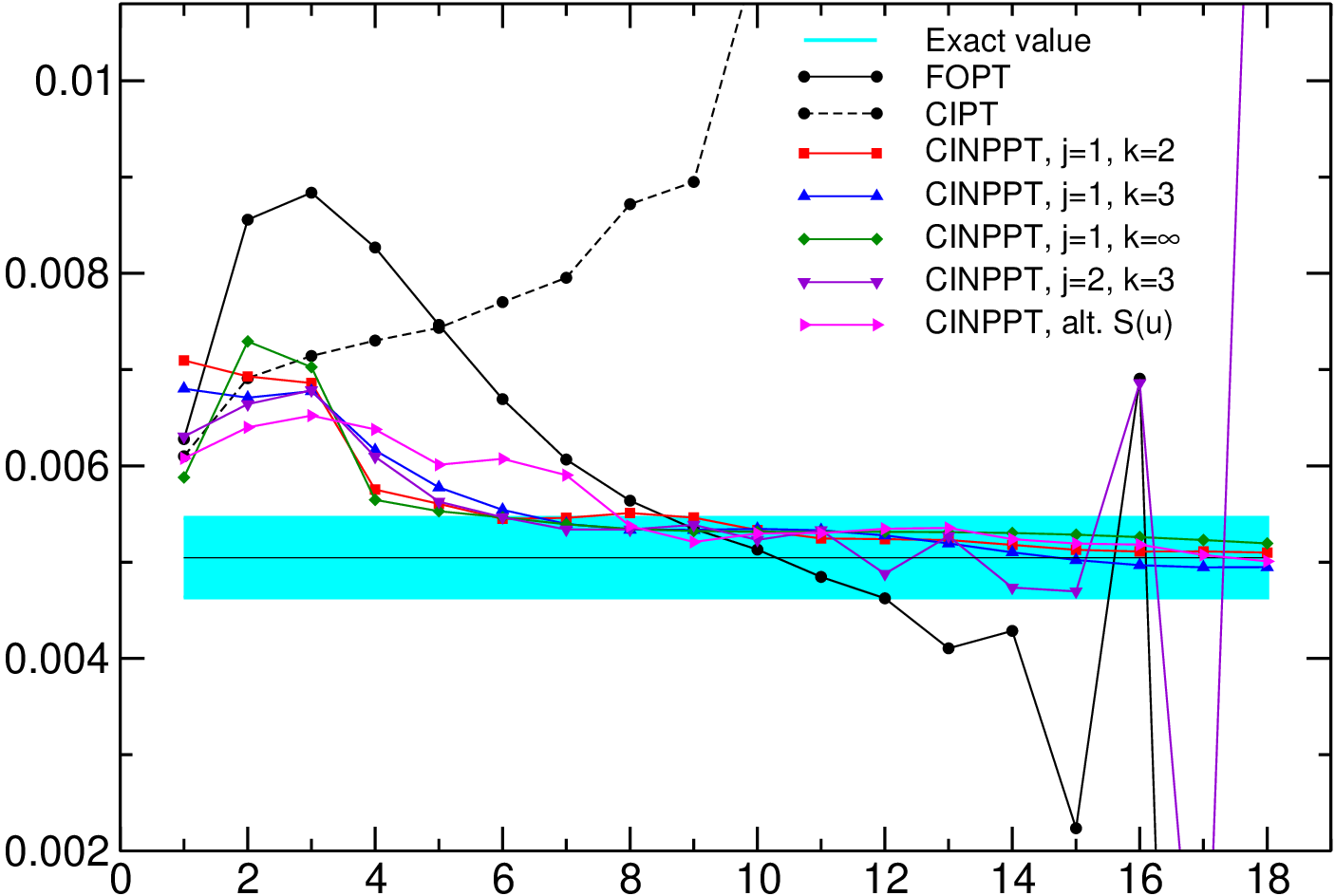}\label{w16RMw}}
\caption{Several CINPPT expansions of the moments shown in Fig. \ref{fig:1}, compared with the standard FOPT and CIPT. The first four nonpower expansions are obtained with the choice (\ref{eq:Sjk}) of the softening factors and several conformal mappings. The  last expansion is obtained using in  (\ref{eq:Bw}) the softening factor $S(u)$ from (\ref{eq:Su}) and the OCM
$\wt w_{12}(u)$. }\label{fig:3}
\end{center}\end{figure*}

The results presented in Fig. \ref{fig:2} and the numerical studies 
performed in the previous works show nevertheless that the optimal 
CINPPT and RGSNPPT may have a slower convergence for models of the 
Adler functions with a residue of the first IR singularity 
significantly smaller than the value it has in the RM. It turns out that the description is less precise
at low orders  also for models where 
this residue is larger than the RM value 
(for such models the standard FOPT and CIPT are both quite poor). Indeed,  the perturbative curves are the same for $N\le 5$ for all models, while by adjusting the residues of the leading singularities one can shift up or down, by a certain amount, the exact values of the moments.

The good convergence of the expansions based on the OCM
for the RM starting from relatively low orders may suggest that this model has a preferred place among models.
Indeed, CINPPT and RGSNPPT have a solid theoretical basis, exploiting simultaneously RG invariance and the known large-order behavior of the expanded function.  However, as mentioned above, there is still a certain arbitrariness in defining these expansions, since the implementation of the singular behavior at the leading branch points is not unique. The preference for the RM  might well be a consequence of the specific choice of the softening factor $S(u)$ given by (\ref{eq:Sjk}), for the OCM defined by  $j=-1$ and $k=2$. In order to reduce the 
possible bias, we must investigate also other expansions, 
with a different  implementation of the threshold behavior. Moreover, as discussed in Sec. \ref{sec:nppt}, for mild singularities  the expansions based on  different conformal mappings are expected to have properties similar to those based on the optimal mapping.  The investigation of a more general class of expansions is the subject of the next subsection.

\subsection{Results obtained with various softening factors and other conformal mappings}
An investigation of CINPPT with different choices of the softening factor $S(u)$ was performed  already  in \cite{CaFi_RJP} for the RM and the particular moment relevant for the $\tau$ hadronic width. For instance, the dominant
behavior (\ref{eq:Bthr})  was implemented by singular factors expressed in terms of the $u$ variable, like in (\ref{eq:Su}), and the leading factors were multiplied  by other functions
analytic in the $u$-complex plane cut along the real axis for $u \ge 2$ and $u \le −1$.
In particular, singularities on an unphysical Riemann sheet,
or placed at $u = 3$ and $u = −2$ were included, the additional factors being expressed
either in the variable $u$ or in the variable $\wt w(u)$.  As reported in \cite{CaFi_RJP}, the results for the  $\tau$ hadronic width are very stable and reproduce well the exact value of the RM for relatively low perturbative orders, of interest for the extraction of $\alpha_s(M_\tau^2)$ from the perturbative calculations available so far. 

 In the present work we consider the class of expansions defined in Sec. \ref{sec:nppt}.  As in \cite{CaFi_Manchester, CaFi2011, AACF,CaSing}, where we investigated the $\tau$ hadronic width,  we adopt besides the OCM $\wt w_{12}(u)$, also the variables $\wt w_{13}(u)$,  $\wt w_{1\infty}(u)$ and $\wt w_{23}(u)$ (some of these conformal mappings have been used also by other authors, see \cite{CaFi2011} for earlier references).  For each expansion variable $\wt w_{jk}(u)$ we chose also a different form of the singularity softening factors $S(u)$, as the simple expression of  $\wt w_{jk}(u)$ given in (\ref{eq:Sjk}).
The only requirement is to reproduce the branch point behavior (\ref{eq:Bthr}). To further enlarge the class, we consider also the softening factor $S(u)$ given by (\ref{eq:Su}).

In Fig. \ref{fig:3} we show  the results obtained with this general class of perturbative expansions. We consider the same moments of the RM as in the previous subsection. 
For simplicity,  we give  only the results obtained in the frame of CINPPT.
The RGSNPPT expansions exhibit a similar behavior.
For comparison we show also the standard CIPT and FOPT. 

The results show that at very low perturbative orders the 
various nonpower expansions are different, but starting from an order $N$ around 5
 they give very similar predictions, 
which  agree also  quite  well with the exact values of the RM moments. One can see that the expansion based on the softening factor (\ref{eq:Su}) gives slightly poorer results
for some moments (for instance the first and the 16th), compared to the expansions based on the softening factors (\ref{eq:Sjk}). We mention that the  softening factor (\ref{eq:Su}) leads to a worse approximation compared to the choice  (\ref{eq:Sjk})  also in the case of the AM. From these results and other numerical tests \cite{CaFi2009,CaFi2011} it follows that the choice of the softening factor as a simple expression (\ref{eq:Sjk}) of the  variable used in the expansion (\ref{eq:Bw}) ensures a good convergence. 

 At larger orders the description is very precise, and
this feature remains stable up to the large-order, $N=18$, 
shown in the figure, and even to larger orders investigated numerically. 
Only the expansion based on the choice $j=2$, $k=3$ starts to exhibit 
oscillations at large-orders (especially for the 16th moment). 
As explained in detail in \cite{CaFi2011}, this behavior is 
due to the effect of the mild  (after singularity softening) 
singularity at $u=-1$, which is still present inside the 
unit disk  $|\wt w_{23}(u)|<1$. This singularity affects
the convergence  of the corresponding power series 
at points $u$ larger than unity, but still small enough such as to 
bring a nonnegligible contribution to the Laplace-Borel integral. 

We conclude that the moments of the RM have  
very stable perturbative expansions in the frame of CINPPT and RGSNPPT, 
for various prescriptions of singularity softening and various 
conformal mappings.  These expansions reproduce the exact moments of 
RM starting from rather low perturbative orders. 
We emphasize that  no assumption  about the magnitude of the 
residues of the singularities is made in defining these expansions. 
\subsection{Results for $s_0 < M_\tau^2$}
In several moment analyses for the extraction of the strong coupling 
and other fundamental parameters of QCD, values of $s_0$ less than $M_\tau^2$, but sufficiently large 
so as to ensure the validity of the perturbation theory, 
have been also employed. For lower values of $s_0$ the convergence 
of the standard perturbative expansions along the circle $|s|=s_0$  is expected to be 
slower due to the fact that the coupling is larger. 
The study of the standard expansions FOPT and CIPT performed in \cite{BBJ} 
was extended to lower values of $s_0$ in \cite{Boito2013}, 
where it was shown that the conclusions of \cite{BBJ} about 
the bad perturbative behavior of some moments and the 
preference for FOPT are still valid for $s_0< M_\tau^2$. 

Here we present the results of our analysis for
the optimal CINPPT  expansions at lower $s_0$.  
As in \cite{Boito2013}, in order to compare the results 
for various $s_0$ we normalize the expansions  
to the exact value of the moment given by the model. 
In Fig. \ref{fig:4} we present the CINPPT expansions for the 
representative moments chosen in this work. 
To keep the figures simple, we do not show now the standard expansions  
(for some of them  see \cite{Boito2013}).  
As expected, the perturbative behavior becomes poorer at lower $s_0$, 
but the extent to which this happens depends very much on the moment. 
On the other hand, the ambiguity of the exact value also increases 
for smaller $s_0$, due to the larger value of $\alpha_s(s_0)$ (we use 
as before, $\alpha_s(M_\tau^2)=0.3186$, which corresponds 
to $\alpha_s(2.5\,\gev^2)=0.3415$ and $\alpha_s(1.5\,\gev^2)=0.4078$). 

For the 2nd, 6th and 13th  moments the perturbative behavior is 
very stable with $s_0$ and within the chosen uncertainty starting 
from low perturbative orders, $N\ge 4$. Therefore, these moments 
are good candidates for moment analyses with lower $s_0$ in the 
framework of CINPPT. The first and the 12th moment 
show stability for $s_0$ down to $2.5\,\gev^2$, 
while at lower $s_0$ the agreement with the 
true value is reached only at higher orders.  
In fact, for these  moments the ambiguity of the 
Borel integral is rather small for the RM. 
Therefore, if we take this uncertainty seriously, 
the perturbative expansions require slightly higher orders, 
$N\ge 6$,  for all $s_0$, to become acceptable. 
Finally, for the 16th moment the CINPPT expansion is 
quite poor at low orders for $s_0=1.5\,\gev^2$, 
but in this case  the ambiguity of the exact value 
is also very large.  
At higher orders the convergence is good in all the cases.

\section{Discussion and conclusions}\label{sec:conc}
In this work we have
investigated several spectral function moments of the massless 
Adler function  in the frame of a new class of ``nonpower" perturbative expansions in QCD, where the 
powers of the coupling are replaced by more 
adequate functions \cite{CaFi1998,CaFi2000,CaFi2001,CaFi2009, CaFi2011, AACF}.  
The new expansions simultaneously  
implement RG summation, either in the ``contour-improved" or in the ``renormalization-group-summed" form,  
and the known location and nature of the first 
singularities of the expanded function in the Borel plane.  
Mathematically, the definition is based on the acceleration
of series convergence by the technique of 
conformal mappings \cite{CiFi} applied in the Borel plane \cite{CaFi1998, CaFi2000, CaFi2001}. 
When reexpanded in powers of $a_s$, 
the new series reproduce order by order the 
perturbative coefficients known from Feynman diagrams. 
On the other hand, they exhibit  a much 
tamer behavior  at larger orders, allowing a more reasonable estimate of the truncation error, which accounts 
for the unknown higher terms in the expansions.

In our earlier works \cite{CaFi2009}-\cite{AACF}, 
the new expansions were used mainly for the extraction of the 
strong coupling from the $\tau$ hadronic width.  
In this work we  go further by employing them 
in a study of other spectral function moments that 
are relevant for the extraction of the strong coupling 
and other QCD parameters from $\tau$ decays. 
Our work is motivated by the recent papers \cite{BBJ,Boito2013}, 
which performed a detailed analysis of the moments in the framework 
of standard CIPT and FOPT. The main aim of our research was to see 
whether the good behavior of CINPPT and RGSNPPT, already established in the case of $\tau$ hadronic width, 
remains valid also for other moments.

In order to assess the quality of various perturbative 
frameworks,  the larger-order pattern of 
the perturbative coefficients of the Adler function must be known. 
Of course, this knowledge is not available and an 
ansatz must be adopted.  
The description of the function in terms of its dominant singularities in the 
Borel plane is a natural choice, 
consistent  with the general principles of analyticity. 
However, a considerable ambiguity still remains  
because, while  the position and nature of the leading 
singularities are known theoretically \cite{Muell, Beneke, BBK,BeJa},  nothing can be said from theory  
about their strengths. 
The recent claims in favor of either CIPT or FOPT are based on  
different views about the magnitude of the 
residues of the leading singularities 
(the IR renormalon at $u=2$ and the  UV renormalon at $u=-1$).
The situation was analysed in detail in \cite{BBJ}, 
where some arguments in favor of a ``reference model", defined in \cite{BeJa}, were put forth. 
Moreover, as discussed in \cite{BBJ}, the reference model 
favors FOPT compared to CIPT.  

Our analysis confirms first the similarity of the ``contour-improved" and  ``renormalization-group-summed" prescriptions, both in the standard form (CIPT and RGSPT) and the nonpower frameworks (CINPPT and RGSNPPT), for all the moments investigated. The essential feature of these  prescriptions is that they sum the large logarithms present in the coefficients  into the running coupling, calculated either numerically (in CIPT) or by explicit expressions (in RGSPT). In the CINPPT and RGSNPPT frameworks the series is further optimized in order to tame the large-order behavior. 

The  results reported in 
Sec. \ref{sec:results}  show that CINPPT and RGSNPPT  
describe very well the spectral function moments of the RM
considered in \cite{BBJ}, including those that are 
poorly described by  the standard expansions, 
FOPT, CIPT and RGSPT.
We have demonstrated a good convergence of CINPPT  for various 
conformal mappings used as  expansion variables after
 softening the leading singularities. The description continues to remain 
good also at lower values of $s_0$, 
within the uncertainties adopted for the true values. 

For the extreme AM defined in \cite{BBJ}, 
where the first IR renormalon is removed by hand, 
the approximation achieved with the nonpower 
expansions defined in Sec. \ref{sec:nppt} 
is less precise for some moments at low orders. This is due to the fact that
 the  expansions are optimally devised
for the physical Adler function, exploiting in a manifest way its first singularities.  
However, they are not optimal for the alternative model, where one of the dominant singularities is absent. 
On the other hand, at higher orders  the nonpower expansions 
have a tame behavior tending to the true values for both models, 
nicely illustrating the theorem of series acceleration by 
conformal mappings \cite{CiFi,CaFi2011}.  

Our analysis shows that  the class of nonpower 
expansions (\ref{eq:cinppt}) and (\ref{eq:rgsnppt}), 
based on different softening factors and different conformal mappings,  
agree among them and with the exact moments of the RM of the Adler function
defined in \cite{BeJa,BBJ} starting from 
rather low perturbative orders, $N=4$ or 5. This may be a coincidence, but may also signal a special place of this model among other  models of the Adler function.  Of course, such a conclusion is not fully rigorous, because the nonpower expansions  contain some arbitrariness in  the implementation of the dominant singular behavior. However, we have investigated several  reasonable  expansions to reduce the bias, and the results are quite stable. Thus, the rapid convergence and the stability of CINPPT and RGSNPPT for all the moments of the RM 
might be an argument in favor of the naturalness of this model. 

In conclusion, the contour-improved nonpower perturbation theory (CINPPT) and the renormalization-group-summed nonpower perturbation 
theory (RGSNPPT) provide a good  perturbative description of 
a large class of  $\tau$ hadronic spectral function  moments, 
including some for which  all the standard expansions fail. In contrast to standard 
perturbation theory, we do not use series in powers of the strong 
coupling, which are mostly chosen for their "simplicity".
  A fundamental merit of our approach is the fact that, to expand 
a singular (Adler, e.g.) function, 
we make 
use of a set of expansion functions possessing singularities that resemble 
those of the expanded function itself.
These expansions also give confidence in a more realistic 
estimate of the truncation error. As a consequence, our 
perturbation expansions  CINPPT and RGSNPPT  provide solid theoretical frameworks 
for the perturbative part in  moment analyses.  
A programme that employs these expansions for the simultaneous 
determination of the strong coupling and other 
parameters of QCD from hadronic $\tau$ decays is of interest for future investigations.

\vspace{-0.3cm}

\subsection*{Acknowledgements} IC acknowledges support from the Ministry of Education under Contracts No. PN 09370102/2009 and No. Idei-PCE 121/2011. The work was supported 
also by the project Nos. LA08015 and LG130131 of the
Ministry of Education of the Czech Republic.

\end{document}